\documentclass[aps,pra,reprint,a4paper,nofootinbib,superscriptaddress,numbers,longbibliography,showpacs,showkeys,floatfix]{revtex4-1}
\usepackage[pdftex]{hyperref,color,graphicx}
\usepackage{amsfonts,amssymb,amsmath,mathrsfs}
\usepackage[separate-uncertainty=true]{siunitx}

\DeclareSIUnit\gauss{G}

\usepackage[english]{babel}
\usepackage[utf8]{inputenc}
\usepackage[T1]{fontenc}
\usepackage{bm}
\usepackage{wasysym}

\usepackage{xspace}

\usepackage{bbm} 
\newcommand\identity{\ensuremath{\mathbbm{1}}}

\DeclareMathOperator{\sinc}{sinc}
\DeclareMathOperator{\module}{mod}

\def\bra#1{\mathinner{\langle{#1}|}}
\def\ket#1{\mathinner{|{#1}\rangle}}

\newif\ifusebibfile
\usebibfiletrue
\newcommand{\figref}[2]{\hyperref[#1]{\ref{#1}(#2)}} 
\newcommand{\encapsulateMath}[1]{\raisebox{0pt}[0pt][0pt]{#1}}

\hypersetup{
    pdftitle = {Creating anomalous Floquet Chern insulators with magnetic quantum walks},
    pdfauthor = {Muhammad Sajid, János Asbóth, Dieter Meschede, Reinhard F. Werner, Andrea Alberti},
	colorlinks=true,linkcolor=blue,citecolor=blue,filecolor=blue,urlcolor=blue,pdfpagemode=UseNone,pdfstartview={XYZ null null 1.00}
  }

 \newcommand{\vect}[1]{\boldsymbol{#1}}    
\newcommand{\position}{\bm{r}}
\newcommand{\step}{n}

\newcommand{\DTQW}{quantum walk}
\newcommand{\BZ}{Brillouin zone}

\newcommand{\TP}{topologically protected\xspace}

\renewcommand\textemdash{\leavevmode\unskip\kern0.8pt\rule[0.19\baselineskip]{8pt}{0.4pt}\kern1pt\ignorespaces}

\begin{document}

\title{Creating anomalous Floquet Chern insulators with magnetic quantum walks}

\author{Muhammad Sajid}
\affiliation{Institut f\"ur Angewandte Physik, Universit\"at Bonn, Wegelerstraße 8, 53115 Bonn, Germany} 

\author{János K. Asbóth}
\affiliation{Institut f\"ur Angewandte Physik, Universit\"at Bonn, Wegelerstraße 8, 53115 Bonn, Germany} 
\affiliation{Institute for Solid State Physics and Optics, Wigner Research Centre for Physics, Hungarian Academy of Sciences, 1525 Budapest P.O.~Box 49, Hungary}

\author{Dieter Meschede}
\affiliation{Institut f\"ur Angewandte Physik, Universit\"at Bonn, Wegelerstraße 8, 53115 Bonn, Germany}

\author{Reinhard F. Werner}
\affiliation{Institut f\"ur Theoretische Physik, Leibniz Universit\"at Hannover, Appelstraße 2, 30167 Hannover, Germany}

\author{Andrea Alberti}
\email{alberti@iap.uni-bonn.de}
\affiliation{Institut f\"ur Angewandte Physik, Universit\"at Bonn, Wegelerstraße 8, 53115 Bonn, Germany} 

\date{\today}

\begin{abstract}

We propose a realistic scheme to construct anomalous Floquet Chern topological insulators using spin-1/2 particles carrying out a discrete-time quantum walk in a two-dimensional lattice.
By Floquet engineering the quantum-walk protocol, an Aharonov-Bohm geometric phase is imprinted onto closed-loop paths in the lattice, thus realizing an abelian gauge field \textemdash the analog of a magnetic flux threading a two-dimensional electron gas.
We show that in the strong field regime, when the flux per plaquette is a sizable fraction of the flux quantum, magnetic quantum walks give rise to nearly flat energy bands featuring nonvanishing Chern numbers.
Furthermore, we find that because of the nonperturbative nature of the periodic driving, a second topological number \textemdash the so-called RLBL invariant \textemdash is necessary to fully characterize the anomalous Floquet topological phases of magnetic quantum walks and to compute the number of topologically protected edge modes expected at the boundaries between different phases.
In the second part of this article, we discuss an implementation of this scheme using neutral atoms in two-dimensional spin-dependent optical lattices, which enables the generation of arbitrary magnetic-field landscapes, including those with sharp boundaries.
The robust atom transport, which is observed along boundaries separating regions of different field strength, reveals the topological character of the Floquet Chern bands.
\end{abstract}

\maketitle

\section{Introduction}
Chern insulators behave as an ordinary band insulator in the bulk, yet exhibit exotic chiral transport in the proximity of boundaries, along which particles can propagate unidirectionally without experiencing backscattering nor dissipation into the bulk.
Such robust transport behavior has its origin in topologically-protected edge modes, which extend all along the length of the insulator.
The existence of \TP edge states is guaranteed by the nontrivial topological structure of the bulk states forming topological bands.
This connection between \TP edge modes and the topological structure of the bulk states is the essence of the \emph{bulk-boundary correspondence} \cite{Hatsugai:1993}.
In a two-dimensional (2D) band insulator, an energy band with a topologically nontrivial structure is characterized by a nonvanishing Chern number \textemdash an invariant that counts the number of topological obstructions to defining a global gauge for the Bloch states of the band \cite{Simon:1983}.

The first Chern insulators to be discovered \cite{thouless1982quantized} are quantum Hall systems, 2D electron gases threaded by a strong magnetic field, which display an extraordinarily robust quantization of their transverse conductance \textemdash a hallmark of \TP edge modes.
Soon thereafter, however, it was realized by Haldane \cite{Haldane1988} that robust chiral transport is not specific to homogeneous magnetic fields,
provided that time-reversal symmetry is broken.
This insight has triggered the quest for topological materials that forego the strong magnetic fields of quantum Hall systems, and yet can conduct charges without dissipation.
Recently this concept has been realized in condensed-matter systems \cite{Chang:2013} and with ultracold atoms trapped in optical lattices \cite{Jotzu:2014,Aidelsburger:2015,Flaschner:2018} in the regime of noninteracting or only weakly interacting particles.

An attractive route to realizing Chern insulators is offered by particles moving in a tight-binding lattice that are subject to a strong, artificial magnetic field \cite{Wu:2012,Scaffidi:2012}.
In fact, in the regime of strong fields, when the flux $\Phi$ threaded through a single plaquette is a sizable fraction of the flux quantum $\Phi_0=h/Q$ ($Q$ is the elementary charge, 
$h$ is the Planck constant), the lattice constant, $a$, becomes comparable with the magnetic length scale, \encapsulateMath{$\ell_B=a\sqrt{\Phi_0/(2\pi\Phi)}$}.
The competition between the two length scales transforms the regular structure of highly degenerate Landau levels, which in the weak-field limit characterize the single particle states, into a fractallike spectrum of energy bands \textemdash the so-called Hofstadter butterfly \cite{Harper:1955,Wannier:1962,Azbel:1964,Hofstadter1976}.
Isolated bands of the Hofstadter spectrum possess nonvanishing Chern numbers, ${C}$, which can generally take large integer values, in stark contrast to the case of Landau levels, which are restricted to $|{C}|=1$.
In addition, for specific ratios $\phi = \Phi/\Phi_0$, the Hofstadter bands are rather flat
and well separated from each other by large energy gaps.

In conventional solid-state materials, attaining the strong field regime requires exorbitantly strong magnetic fields of several thousand 
teslas.
More favorable conditions are achieved using artificially engineered superlattices \cite{Albrecht:2001} and moiré superlattices made of graphene on a semiconductor substrate \cite{Dean:2013}, where the required field strength is reduced by several orders of magnitude.
Yet, very high magnetic fields remain needed.
To avoid dealing with such high fields altogether, a number of proposals have been put forward \cite{Dalibard:2011,Goldman:2014,Aidelsburger:2017} aiming at recreating artificially the effect of magnetic fields through engineered Aharonov-Bohm geometric phases.

For neutral atoms in optical lattices, the standard approach to create such an artificial gauge field relies on photon-assisted tunneling.
This can be realized either using an additional 
optical dressing field \cite{Jaksch:2003,Ruostekoski:2002,Mueller:2004,Gerbier:2010,Cooper:2012} or by shaking the lattice itself \cite{Kolovsky2011,Tarallo:2012,Creffield:2016}.
Through this process, the hopping terms of the tight-binding Hamiltonian acquire a complex phase factor \textemdash the so-called Peierls phase \cite{Peierls:1933,Luttinger:1951};
integrating these phases along a closed contour yields (in units of $\Phi_0/2\pi$) the flux of the artificial gauge field enclosed therein.
Following this approach, significant experimental progress has been made over the past ten years, first demonstrating complex hopping amplitudes in one-dimensional (1D) optical lattices  \cite{Alberti:2010,JimnezGarcia:2012,Struck:2012} and, subsequently, strong artificial magnetic fields in 2D optical lattices \cite{Aidelsburger:2013,Miyake:2013,Jotzu:2014,Aidelsburger:2015}.
However, because the kinetic energy in a shallow lattice is of order of the recoil energy $E_R=\hbar^2/(m\hspace{0.3pt}a^2)$ ($a$ is the lattice constant and $m$ is the atomic mass), ultracold-atom experiments in optical lattices must be conducted at very low energy scales, corresponding to few nK, which are generally difficult to reach.
In addition, low kinetic energies imply small hopping terms ($<h\times\SI{1}{\kilo\hertz}$) and, correspondingly, long evolution times, during which heating \cite{Aidelsburger:2013,Miyake:2013,Lellouch:2017,Reitter:2017} and other decoherence mechanisms can have a detrimental effect on the coherent evolution of the system.
Increasing the kinetic energy by using light atomic species (e.g., lithium) and, possibly, opting for subwavelength lattice constants \cite{Yi:2008,Nascimbene:2015,Wang:2018} have been identified as effective measures to increase the kinetic energy scale, thus curbing the technical challenges faced by experimental implementations.

In this article, we map out a different route to flat-band Chern insulators, which uses discrete-time quantum walks of ultracold atoms on a square lattice to create artificial gauge fields.
The general idea of this work draws inspiration from an early proposal by Sørensen \emph{et al.}\ \cite{Sorensen:2005} and  a subsequent development by Creffield \emph{et al.}\ \cite{Creffield:2014}, where the desired time evolution is constructed from a periodic sequence of unitary operations that are applied at discrete time intervals.
In a discrete-time quantum walk, in fact, the various degrees of freedom (e.g., motion in the $x$- and $y$-direction) evolve at different times.
Following this idea, we show that Peierls phases can be imprinted onto the walker's wavefunction at a time subsequent to its motional dynamics.
This provides a great amount of flexibility, which can be used to create arbitrary magnetic-field landscapes, including boundaries between different magnetic domains.
By computing the Chern numbers of the energy bands, and studying the excitation of \TP edge modes at the boundaries of magnetic domains, we show that \emph{magnetic quantum walks}, realized by Floquet engineering of the Peierls phases, behave like a Floquet Chern insulator.

In a magnetic quantum walk, however, the Chern numbers alone do not fully capture the bulk topology.
Quantum walks are in fact periodically driven systems characterized by large driving amplitudes and small modulation frequencies, where the periodic driving cannot be treated using perturbative approaches \cite{Rudner2013,Asboth:2015}.
As a result, magnetic quantum walks are anomalous Floquet topological insulators, hosting  topologically protected edge states that cannot be predicted only using Chern numbers.
The number of such anomalous edge modes is determined by a winding number of the periodic protocol, which can be associated with each quasienergy gap; we call this topological number RLBL invariant after Rudner \emph{et al.} \cite{Rudner2013}.
Such anomalous edge states have been studied in an experimental proposal for ultracold atoms \cite{Groh2016}, and recently observed in photonic systems \cite{Maczewsky:2017,Wang:2018a,Chen:2018}.
In a magnetic quantum walk, both the Chern numbers and the RLBL invariant play a role in determining the bulk topology and edge states.
For the experimental realization of magnetic quantum walks, we propose to use fast shift operations in deep optical lattices, which displace atoms by an integer number of lattice sites, depending on their internal spin state.
Thereby, one can entirely forego the slow dynamics of approaches based on photon-assisted tunneling, and yet achieve delocalization of matter waves.
Peierls phases are controlled by applying onto the atoms a spin-dependent potential for a short duration of time $\tau$, which can be realized by illuminating the atoms with a suitably designed intensity pattern.
By tuning $\tau$ to certain magic values and by taking advantage of Floquet engineering, we show that motional excitations of atoms and off-resonant photon scattering can be strongly suppressed. 
Moreover, shifting atoms by several lattice sites \cite{Robens:2015} allows magnetic quantum walks to be realized on a superlattice with an augmented lattice constant, without significantly affecting the time required for a single step of the walk.  
The larger lattice constant results in an effective increase of the optical resolution of the imaging system imprinting the Peierls phases, which in turn can be used to lessen the demands on the numerical aperture of objective lens and the pointing stability of laser beams.

\section{Discrete-time quantum walk in an artificial magnetic field} \label{section:2DQW}

\begin{figure*}[t]
 \includegraphics[width=1\textwidth]{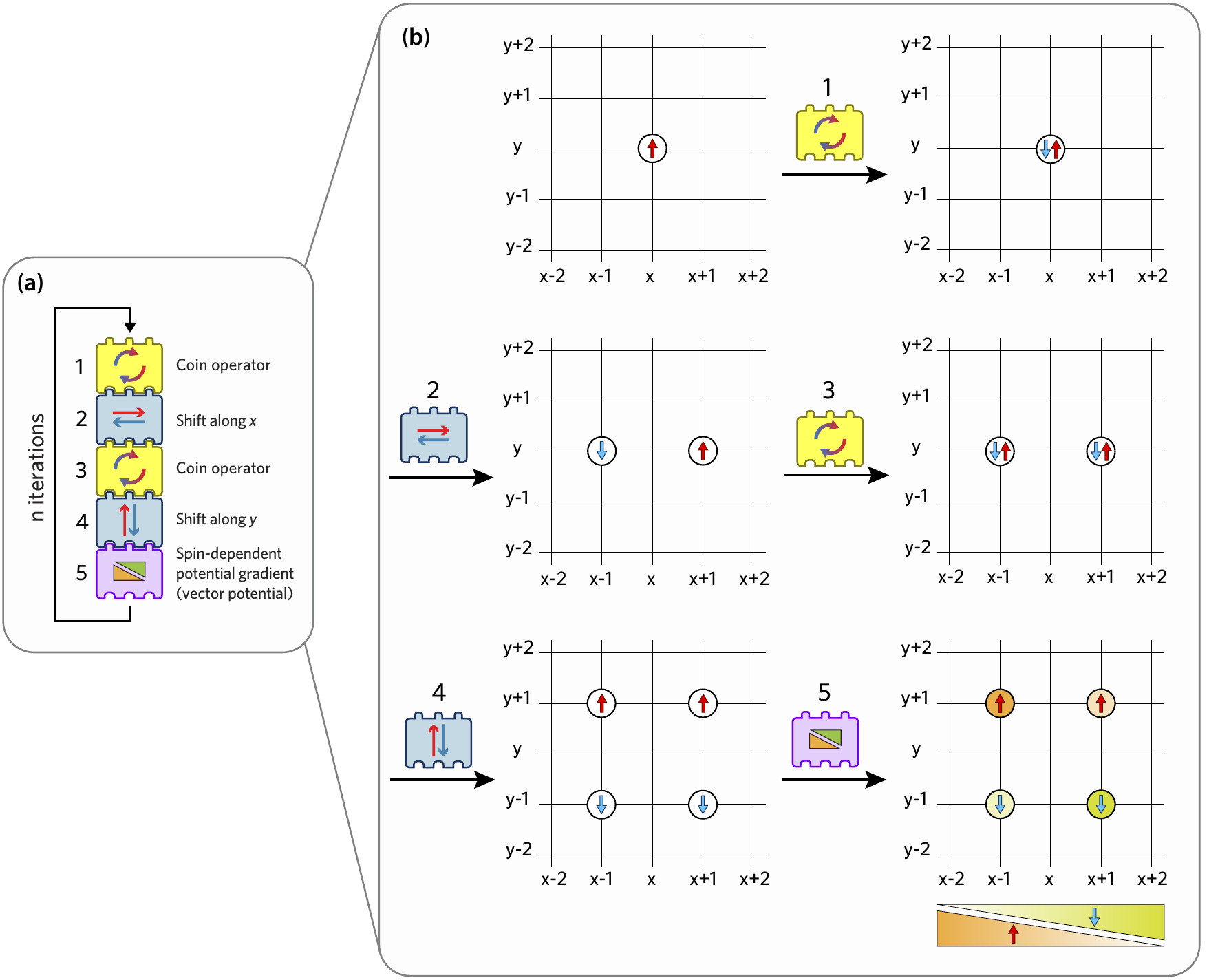}
 \caption{Magnetic \DTQW{}s on a 2D lattice.
(a) Left: block representation of the unitary operators constituting a single time step of the evolution, \encapsulateMath{$\hat{W}$}, as defined in Eq.~(\ref{eq:main_1}).
(b) A single time step of a walker prepared initially in a spin-up state, represented on a square lattice, with the arrows denoting the two spin states. To simulate a homogeneous magnetic field $B$, with artificial vector potential $\vect{A}=(0,Bx,0)$, a spin-dependent linear potential gradient is stroboscopically applied along the $x$-axis, imprinting onto the walker's wavefunction a spin-dependent linear phase gradient (block 5).
}
\label{fig:Peierls Phase}
\end{figure*}

We consider a single particle, also called a walker, that moves in discrete steps on a square lattice.
Its position states, $\ket{\position}$, are labeled by the lattice coordinates, $\position = (x,y) \in \mathbb{Z}^2$.
Similarly to a spin-1/2 particle, the walker possesses
two internal states,
$\ket{\uparrow}$ and $\ket{\downarrow}$,
which condition the  motion of the particle by deciding the shift direction.
Moreover, for convenience, we consider in this section dimensionless units, assuming the lattice constant, $a$, the artificial charge, $Q$, the single-step duration, $T$, and the reduced Planck constant, $\hbar$, to be all equal to 1.

The protocol of a magnetic quantum walk is defined by the repeated application 
of the time-step operator,
\begin{align}
\label{eq:main_1}
\hat{W} = \hat{F} \, \hat{S}_y \, \hat{C} \, \hat{S}_x \, \hat{C},
\end{align}
consisting of a series of unitary operators, which comprise: the coin, \encapsulateMath{$\hat{C}$}, the shifts along the $x$- and $y$-axes, \encapsulateMath{$\hat{S}_x$}, \encapsulateMath{$\hat{S}_y$}, and the magnetic-field operator \encapsulateMath{$\hat{F}$}.
The effect of each operator is illustrated in Fig.~\ref{fig:Peierls Phase}, and described in more detail below.

The coin is simply a rotation of the walker's spin state. 
It is independent of the position and is represented by the Hadamard-like operator \cite{DiFranco2011}:
\begin{equation} \label{eq:coin} \hat{C} = \exp(-i\hat{\sigma}_y {\pi}/{4}) = \frac{1}{\sqrt{2}} \begin{pmatrix} 1 & -1 \\ 1 & \ \ 1 \end{pmatrix},
\end{equation}
with $\hat{\sigma}_i$ denoting the $i$-th Pauli matrix.

\begin{figure*}[t]
\centering
\includegraphics[width=\textwidth]{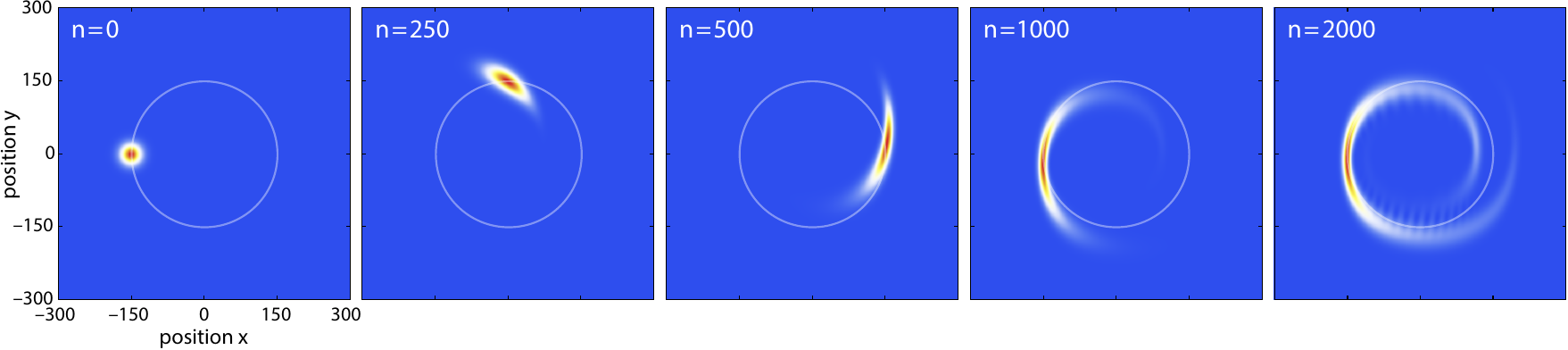}
\caption{Time evolution of a 
walker 
in a weak magnetic field, $\phi=1/1200$.
The walker is initially prepared in a 
Gaussian wave packet with a root-mean-square width of 15 sites and a well-defined momentum state chosen at a distance $\delta{k}=\pi/4$ from one of the four inequivalent Dirac points [see Fig.~\figref{fig:quasienergy_spectrum_bulk}{a}]. Its initial spin state is oriented such that only energy states in the upper Dirac cone are occupied.
The color scale represents the spatial probability distribution
after $n$ steps of the walk, with dark red indicating a high, white an intermediate and blue a vanishing probability.
According to semiclassical equations (see main text), we expect a cyclotron circular orbit of 150 sites radius (white circle) and a revolution period of 942 steps.
An animation of the time evolution is provided in the Supplemental Material \cite{VideoWeakFields}.}
\label{fig:weakfields}
\end{figure*}

The shift operators are spin-dependent spatial translations of the walker by one lattice site:
\begin{equation} \label{eq:shift-x-y}
\hat{S}_d = \sum_{\position} \ket{\position+\vect{e}_d}\bra{\position} \otimes \ket{\uparrow}\bra{\uparrow} + \ket{\position-\vect{e}_d}\bra{\position} \otimes \ket{\downarrow}\bra{\downarrow},
\end{equation} 
with $\vect{e}_d$ representing the unit lattice vector in the $d$-direction ($d\in \{x,y\}$).
With this definition, spin-up particles are shifted rightward by \encapsulateMath{$\hat{S}_x$} and upward by \encapsulateMath{$\hat{S}_y$}, while spin-down particles are shifted 
in the opposite directions.
Owing to their invariance under lattice translations, the two shift operators can be expressed more conveniently as
\begin{equation}
	\label{eq:shiftoperator}
	\hat{S}_d=\exp(-i \hat{\sigma}_z  \hat{k}_d) =  \begin{pmatrix} e^{-i \hat{k}_d} & 0 \\ 0 & e^{i \hat{k}_d} \end{pmatrix},
\end{equation}
where \encapsulateMath{$\hat{k}$} represents the quasimomentum operator associated with the $d$-direction of the square lattice, taking values in the interval $[-\pi,\pi]$, with the endpoints identified.

To simulate the effect of a magnetic field coupled to the walker, spin-dependent phases are stroboscopically imprinted onto the walker's wavefunction by the so-called magnetic-field operator, \encapsulateMath{$\hat{F}$}.
These phases act like the Peierls phases in the Hofstadter Hamiltonian \cite{Peierls:1933,Luttinger:1951}, and are determined by the vector potential associated with the artificial magnetic field.
Magnetic field operators simulating an arbitrary vector potential are discussed in Appendix~\ref{artificial_vector_potential}.
For a vector potential in the Landau gauge, $\boldsymbol{A} = (0,Bx,0)$, resulting in a homogeneous magnetic field $B$, the magnetic-field operator reads as
\begin{equation}
\label{eq:spin-dependent_phase_shift}
\hat{F} = \exp(i \hat{\sigma}_z B \hat{x})= \begin{pmatrix} e^{i B \hat{x}} & 0 \\ 0 & e^{-i B \hat{x}} \end{pmatrix},
\end{equation}
with $\hat{x}$ being the lattice position operator along the $x$-axis.  As a result of this operator, the walker's wave function acquires on a closed-loop path an Aharonov-Bohm phase equal to $B = 2\pi \phi$, times the number of plaquettes enclosed by the path itself, with $\phi$ being the flux per plaquette.
Shifting the magnetic flux by an integer number of flux quanta
leaves 
the magnetic-field operator in Eq.~(\ref{eq:spin-dependent_phase_shift})
unchanged.

The physical mechanism creating an artificial gauge field can be understood by rewriting the product of the last two operators in Eq.~(\ref{eq:main_1}) as:
\begin{equation}
\hat{F}\hat{S}_y = \exp[-i\hat{\sigma}_z(\hat{k}_y-\hat{A}_y)],
\end{equation}
where $A_y$ is the $y$-component of the engineered vector potential $\vect{A}$.
In this form, one can clearly recognize \cite{Yalcinkaya2015} 
that the quasimomentum operator is shifted by an amount proportional to the vector potential, in a similar fashion as minimal coupling in classical electromagnetism.
Equivalently, the magnetic-field operator can be thought of as a way to ensure discrete local gauge invariance \cite{Bialynicki-Birula:1994,Arnault2016b,Marquez-Martin:2018,Arnault:2019,Cedzich:2019} of the discrete-time quantum-walk protocol. 

A further insight into the effect of the magnetic field operator is obtained by considering the dynamics of a magnetic quantum walk in the weak-field regime, $\phi\ll 1$, when the magnetic length scale is much larger than the lattice constant, $\ell_B\gg a$.
In these conditions, semiclassical equations based on the long-wavelength approximation can be used to describe the dynamics of a wave packet with a narrow quasimomentum spread \cite{Luttinger:1951}.
Figure~\ref{fig:weakfields} shows the simulated evolution of a walker in a weak magnetic flux, $\phi=1/1200$, with the  quasimomentum of the initial wave packet prepared at a small distance, $\delta k$, from one of the Dirac points of the walker's spectrum, which is discussed in detail in Sec.~\ref{subsection:bulk_spectrum}.
In the vicinity of a Dirac point, the walker mimics the behavior of a massless Dirac fermion \cite{Kim:2017}, moving with a constant velocity modulus and subject to a uniform magnetic field.
The resulting Lorentz force \cite{Luttinger:1951} deflects the wave packet's quasimomentum on a circular trajectory enclosing the Dirac point.
Knowing that the walker's velocity in the vicinity of a Dirac point is $a/T=1$ (see Appendix~\ref{app:dirac_cones}), one can directly show that the wave packet follows in real space a circular trajectory with radius  \encapsulateMath{$\delta{k}/(2\pi\phi)$} and period \encapsulateMath{$\delta{k}/\phi$}, as shown in Fig.~\ref{fig:weakfields}.

The weak-field regime also allows us to obtain a first intuition into the topological properties of magnetic quantum walks, focussing on low energies where the dispersion relation resembles Dirac cones.
In this regime, the walker behaves like a massless Dirac fermion coupled to a magnetic field, whose topological transport properties have been extensively studied in recent years in relation to graphene physics \cite{Gusynin2005,Novoselov2005,Arnault2016}.
The Dirac spectrum splits into a series of highly degenerate Landau levels (see also Fig.~\ref{fig:Hofstadter_butterfly}), with each level characterized by a Chern number $C=2$ \cite{Hatsugai:2006}.
This value originates from the fact that Dirac cones appear in pairs in a lattice system (fermion doubling, see Appendix~\ref{app:sublattice_symmetries}).

While the weak-field limit is well understood within a semiclassical approach, as discussed above, the strong-field regime,  in which the magnetic length scale $\ell_B$ is comparable with the lattice constant $a$, requires a different approach, to which the rest of this work is devoted.

\subsection{Quasienergy spectrum: the Floquet--Hofstadter butterfly} \label{subsection:bulk_spectrum}

The protocol driving the walker is invariant under discrete time translations by an integer number of steps.
 Based on this invariance, we can use
an effective static Hamiltonian \encapsulateMath{$\hat{H}_\text{eff}$} (also known as Floquet Hamiltonian) to describe the time-evolved state of the walker, $\ket{\psi_n}$, after $n$ discrete steps:
\begin{equation}
	\label{eq:Heff}
	\ket{\Psi_{\step}} =  \hat{W}^n \ket{\Psi_0}=\exp(-i n \hat{H}_\text{eff}) \ket{\Psi_{0}}.
\end{equation}
The effective Hamiltonian is simply
defined as the complex logarithm of the time-step operator,
\encapsulateMath{$\hat{H}_\text{eff}=i\log(\hat{W})$}.
Its eigenvalues are called quasienergies, and are only defined up to an integer multiple of $2\pi$, reflecting the freedom in choosing the branch cut of the logarithm. 
In this work, we fix the branch cut (unless otherwise specified) along the negative real axis, so that quasienergies are represented in the interval $[-\pi,\pi]$, with the endpoints identified.
We call this interval the Floquet zone, by analogy with the Brillouin zone used to represent quasimomentum in a translationally invariant lattice system.	

\begin{figure}[t]
 \centering
 \includegraphics[width=\columnwidth]{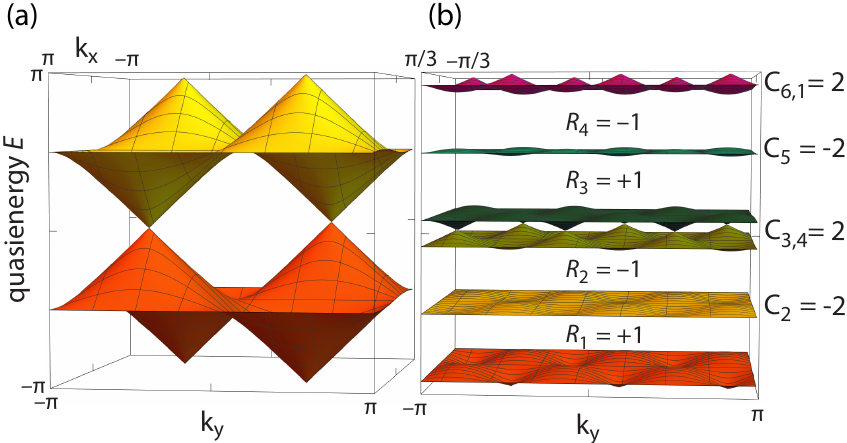}
 \caption{Quasienergy spectrum of magnetic quantum walks as a function of quasimomentum $(k_x,k_y)$. (a) In a zero magnetic field, the spectrum 
 presents two bands, touching each other at four inequivalent Dirac points.
The number of bands reflects the spin multiplicity.
(b) For a high magnetic field, $\phi=1/3$, the original gapless spectrum splits into nearly flat quasienergy bands, separated in some cases by a large gap.
The Chern numbers of isolated bands are denoted by $C_i$ with $i$ denoting the respective band index, or set of indices in the case of two bands touching.
The Floquet invariants, $R_i$, associated with the $i$-th quasienergy gap are also shown.
} \label{fig:quasienergy_spectrum_bulk}
\end{figure}

The spatial period of a magnetic quantum walk is bigger than the lattice constant, since the magnetic-field operator, \encapsulateMath{$\hat{F}$}, breaks the translational symmetry of the underlying lattice.
For our choice of the Landau gauge, 
and for a magnetic field $B$ with flux per plaquette $\phi=p/q$, where $p$ and $q$ are coprime integers, the smallest unit repeating itself is constituted by $q \times 1$ plaquettes of the square lattice, which we call the magnetic unit cell.
In reciprocal space, the magnetic unit cell corresponds to a \BZ{}{} that is shrunken in the $k_x$-direction by a factor $q$, i.e., with 
$k_x$ defined in the interval $[-\pi/q,\pi/q]$, $k_y$ in $[-\pi,\pi]$, and the endpoints of these intervals identified.

The quasienergy spectrum of the walker is shown in Fig.~\ref{fig:quasienergy_spectrum_bulk}, comparing the situation of no flux with that of a high magnetic flux, $\phi=1/3$. 
For a vanishing magnetic field [see Fig.~\figref{fig:quasienergy_spectrum_bulk}{a}], the spectrum presents two quasienergy bands, touching each other at four inequivalent Dirac points, occurring in pairs at quasienergy $E=0$ and $E=\pi$.
These band touching points are topologically protected, since  in their vicinity  the effective Hamiltonian has a Rashba type of spin-orbit coupling, $\hat{H}_\text{eff} \approx \mp \delta k_x \hat{\sigma}_y + \delta k_y \hat{\sigma}_z$, carrying a nonzero topological charge; for more detail, see Appendices~\ref{app:dirac_cones} and \ref{app:top_charges}.
For a high magnetic field [see Fig.~\figref{fig:quasienergy_spectrum_bulk}{b}], the two original bands split, we find $2\times q$ quasienergy bands, which are rather flat and, in some cases, well separated by a large quasienergy gap.
For odd $q$, Dirac points are present at quasienergies $0$ and $\pi$, $q$ times as many as in the zero-field case owing to the larger magnetic unit cell \cite{rhim2012self}.

In both cases, with and without magnetic field, the spectrum is mirror symmetric with respect to the plane $E=0$ and $E=\pi$ as a result of chiral symmetry, which is discussed in detail in Appendix~\ref{app:chiral_symmetry}.
Chiral symmetry plays a crucial role for ensuring the stability of the Dirac points, as discussed in Appendix~\ref{app:top_charges}.
In addition to chiral symmetry, the quantum walk has sublattice symmetries that impose additional constraints on the quasienergy dispersion relation, as shown in Appendix \ref{app:sublattice_symmetries}.

\begin{figure}[t]
 \centering
 \includegraphics[width=\columnwidth]{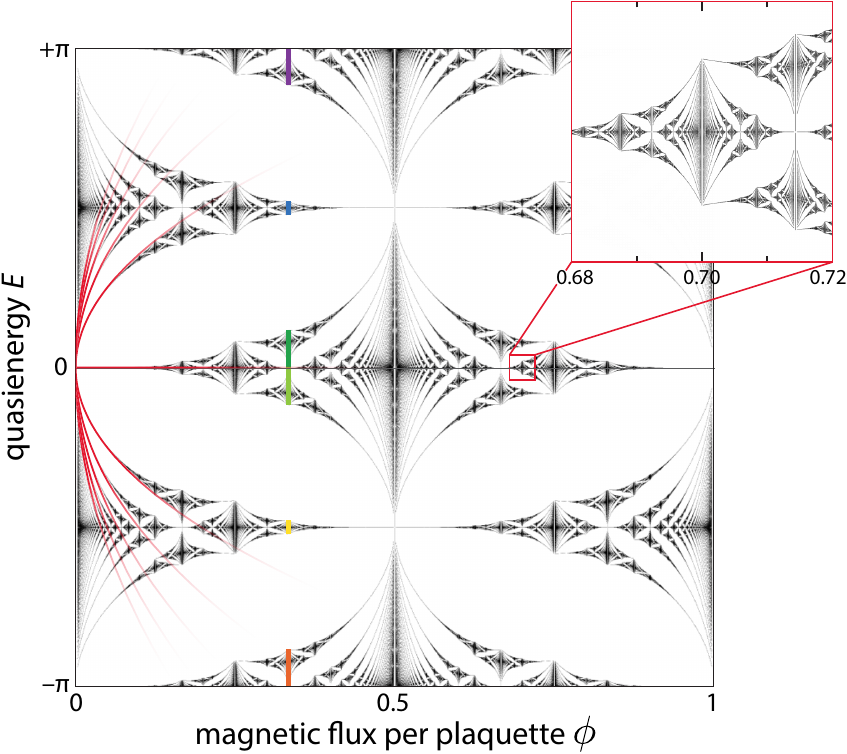}
 \caption{Floquet--Hofstadter butterfly showing the quasienergy spectrum of magnetic quantum walks as a function of the magnetic flux $\phi$.
For each rational value of the flux $\phi=p/q$, there are $2q$ bands reflecting the spin multiplicity and the size of the magnetic unit cell.
The six colored vertical lines at $\phi=1/3$ correspond to the quasienergy bands shown in Fig.~\figref{fig:quasienergy_spectrum_bulk}{b}.
On the left-hand side, the overlaid parabolic red lines, $E=\sqrt{4\pi\, n\, \phi}$, indicate the Landau levels for a massless Dirac particle in the weak-field limit \cite{rhim2012self}. Inset: a detail of the butterfly, illustrating its self-similarity.
}
 \label{fig:Hofstadter_butterfly}
\end{figure}

The quasienergy spectrum as a function of the magnetic flux is an intricate fractallike structure  shown in Fig~\ref{fig:Hofstadter_butterfly}.
The  spectrum bears a close resemblance to the celebrated Hofstadter butterfly \cite{Azbel:1964,Hofstadter1976},
describing the energy levels of a spinless charged particle in a tight-binding lattice under the influence of a uniform magnetic field.
There is, however, one important difference that the butterfly here exhibits a periodicity not only in $\phi$, but also in quasienergy $E$.

\subsection{Topological invariants}
\label{sec:top_invariants}
The Chern number is a topological invariant that can be assigned in a 2D lattice system to any set of bands separated from all others by energy gaps.
For a non-degenerate band, it is defined as the integral over the \BZ{}{} of the Berry curvature associated with the states in that band \cite{ChernNumber}.
A non-vanishing Chern number represents an obstruction to defining a smooth global gauge over the whole magnetic Brillouin zone for the Bloch wavefunctions of that particular set of bands.

For a magnetic quantum walk, a Chern number, $C$, can be assigned to each set of quasienergy bands of the effective Hamiltonian, in the same way as for static 2D band insulators.
Following this idea, we calculate the Chern number of each set of isolated bands using an efficient numerical algorithm  \cite{Fukui2005}, which counts in the \BZ{}{} the number of vortices of the determinant of the 
Wilczek--Zee connection for that particular set of bands.
The  Chern numbers obtained for $\phi=1/3$ are shown in Fig.~\figref{fig:quasienergy_spectrum_bulk}{b} next to the quasienergy bands.
As expected, based on chiral symmetry (see Appendix~\ref{app:chiral_symmetry}), 
the set of Chern numbers is mirror symmetric with respect to the energy planes $E=0$ and $E=\pi$.
Moreover, because of sublattice symmetry (see Appendix~\ref{app:sublattice_symmetries}), the Chern number of a set of bands around quasienergy $E$ is the same as that of the bands around $E\pm\pi$.

Chern numbers alone, however, do not provide a full characterization of the topological phases of magnetic quantum walks.
In such an anomalous Floquet topological insulator, this can be done using the so-called RLBL invariant \cite{Rudner2013}, which can be computed for each quasienergy gap, and is defined as a three-dimensional winding number of the ``periodized'' time-step operator (winding along quasimomenta $k_x$, $k_y$, and time $t$).
It represents a topological property of the periodic quantum-walk protocol \cite{Asboth:2015}, which is not captured by the effective Hamiltonian $\hat{H}_\text{eff}$.

The RLBL invariant as defined in Ref.~\cite{Rudner2013} is an integer assigned to the quasienergy gap comprising the endpoint of the Floquet zone.
For our choice of the branch cut of the logarithm defining $\hat{H}_\text{eff}$, this corresponds to quasienergy $E=\pi$.
However, changing the choice of the branch cut, it is possible to calculate the RLBL invariant for all gaps of the quasienergy spectrum.
These gap invariants, $R_i$, provide a complete classification of the topological phases of a magnetic quantum walk.
From these gap invariants it is possible to obtain the Chern numbers while the converse is not true \cite{Rudner2013}:
In fact, the difference between two gap invariants corresponding to different quasienergy gaps is equal to the sum of the Chern numbers of all bands lying between the two quasienergy gaps.

We calculate the RLBL gap invariants, $R_i$, for all gaps of the quasienergy spectrum with magnetic flux $\phi=1/3$.
To that end, instead of evaluating the rather involved three-dimensional winding number of the ``periodized'' time-step operator, as done in previous literature \cite{Rudner2013,Asboth:2015}, we here employ a 
simpler method \cite{Asboth2017}, measuring the spectral flow induced by a fictitious magnetic field, which is added on top of the artificial magnetic field $B$.
We show in Fig.~\figref{fig:quasienergy_spectrum_bulk}{b} the obtained values of $R$, and provide the details of their calculation in Appendix~\ref{sec:spectral_flow}.
As expected, the difference between two gap invariants associated with adjacent quasienergy gaps is equal to the Chern number of the band (or set of bands) in between the selected quasienergy gaps.
In addition, we observe that because of chiral symmetry, the RLBL invariants are mirror antisymmetric with respect to the energy planes $E=0$ and $E=\pi$, meaning that mirror pairs of RLBL invariants have the same value but opposite sign.
Sublattice symmetry ensures in addition that the RLBL invariant of a gap containing quasienergy $E$ is the same as that of a gap at quasienergy $E\pm\pi$.

\subsection{Topologically protected edge states}\label{sec.spatial boundaries}

For a topological insulator with boundaries, a bulk-boundary correspondence applies, relating the gap invariants, which characterize the topology of bulk 
bands, to the existence of edge states localized at the boundaries \cite{Hatsugai:1993,Asboth}.
According to this correspondence, for a boundary between two bulks (say, bulk A and bulk B), at any energy that is in a gap of both bulk spectra, the minimal number of edge states corresponds to the difference of the gap invariants associated with the two gaps (i.e., $R_A-R_B$).
These edge states are topologically protected and are a hallmark of topological insulators.

In a topological insulator, the boundary can just be the physical edge of the sample.
Although boundaries of this type can be created in a magnetic quantum walk, simulating the effect of an edge potential or of a vacuum state \cite{Asboth:2015}, we here focus on a different scenario, where the boundary separates regions with different magnetic fluxes.
We are in particular interested in the situation of a high magnetic flux, say, $\phi=1/3$, that across the boundary, on the length scale of a single lattice site, inverts its sign to $\phi=-1/3$.
In terms of conventional solid-state materials, this situation corresponds to a bulk region with a homogeneous field of the order of $\SI{e5}{\tesla}$, inverting its sign abruptly over just a few angstroms at the interface with a second region of opposite magnetic flux.
These are exorbitantly large fields and fields gradients, which can only be experimentally realized in artificial materials such as magnetic quantum walks.
Related experiments in solid-state systems have investigated electronic transport at an interface between regions with opposite magnetic flux, however, not in the regime of strong magnetic fields.
In these experiments, so-called snake states have been realized, either by inverting the magnetic field direction in a 2D electron gas \cite{Weiss1995}, or by changing the type of charge carriers using graphene p-n junctions \cite{Rickhaus2015, Liu2015}.

\begin{figure}[t]
 \centering
 \includegraphics[width=\columnwidth]{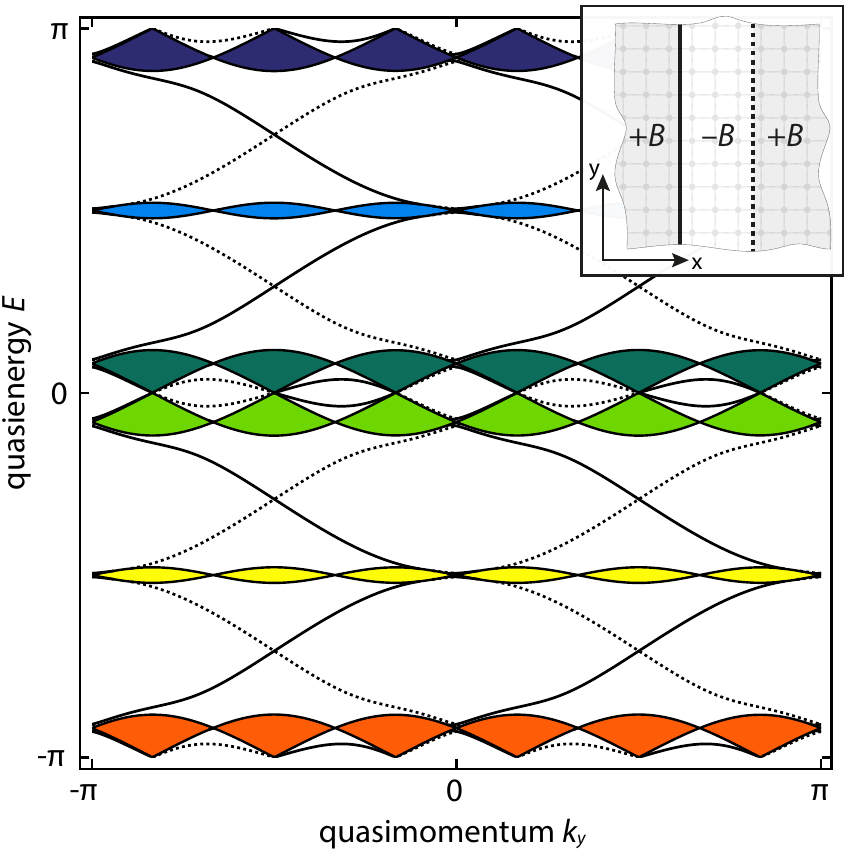}
 \caption{Quasienergy spectrum  as a function of quasimomentum $k_y$, computed for a quantum walk with the magnetic field inverted in a central stripe.
Inset: Schematic representation (not to scale) of the magnetic-field landscape, with $\phi=1/3$ outside the stripe and $\phi=-1/3$ inside.
Main figure: The filled bands correspond to bulk states, whereas the midgap energy branches to \TP edge modes, with the solid (dotted) lines denoting the left (right) edge of the stripe.
The net number of \TP edge modes (per edge) matches the difference between the gap topological invariants [cf.\ Fig.~\figref{fig:quasienergy_spectrum_bulk}{b}] of the two different topological domains.
The spectrum is computed under realistic conditions, based on the Floquet phase-imprinting scheme detailed in Sec.~\ref{sec:hommagfield}, assuming periodical boundary conditions along the $x$-direction, and considering $60$ sites in total, with the left edge at $x=15$ and the right one at $x=45$.}
\label{fig:bulk_and_edge_spectrum} 
\end{figure}

\begin{figure*}[t]
\centering
\includegraphics[width=\textwidth]{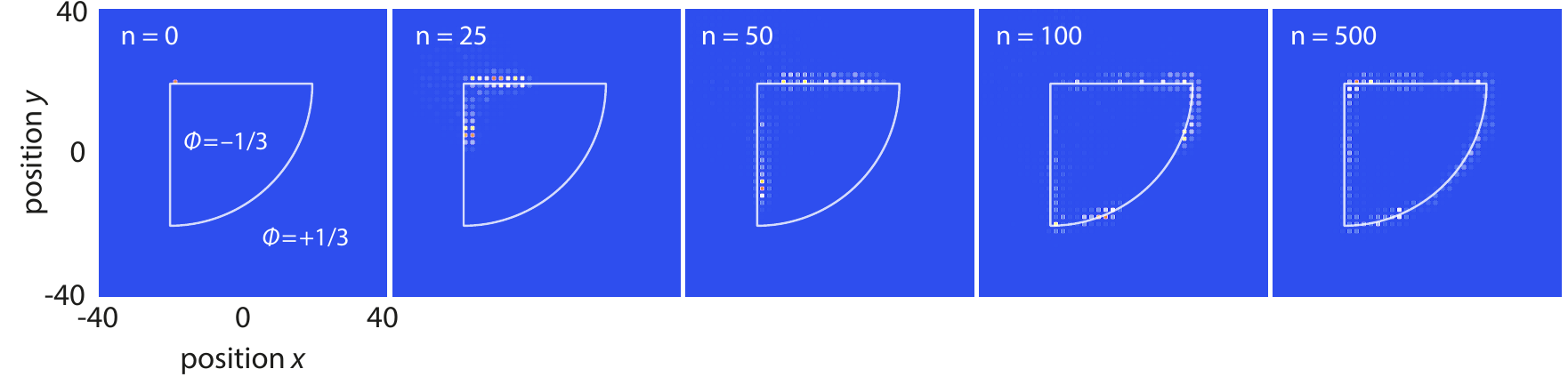}
\caption{Time evolution of a walker that is initially prepared in a single site close to the boundary separating two distinct, topologically nontrivial regions.
The boundary (overlaid white curve) has the shape of a quarter of a circle with a radius of 40 sites.
The magnetic flux $\phi$ is $- 1/3$ in the inner region, and $1/3$ in the outer one.
The color scale, similarly as in Fig.~\ref{fig:weakfields}, represents the spatial probability distribution after $n$ steps of the walk. 
The simulations are carried out numerically assuming realistic conditions, based on the Floquet phase-imprinting scheme detailed in Sec.~\ref{sec:hommagfield}, using 
a simple lattice with a relative shift of $0.5$ (cf.~Fig.~\ref{fig:gapwidthVsshift}) with respect to the sawtooth intensity profile employed to imprint the Peierls phases.
An animation showing the time evolution of the walker is provided in the Supplemental Material \cite{VideoBoundary}.
}
\label{fig:evolution_along_quarter_circular_boundary}
\end{figure*}

The salient feature of \TP edge states is that their energies form continuous branches of the dispersion relation, connecting adjacent energy bands through the bulk gaps.
These branches of the dispersion relation,  
called hereafter the \TP edge modes, are responsible for the robust, quantized, directional transport of charges characterizing topological insulators.
In a magnetic quantum walk, \TP edge modes can be clearly identified by considering the quasienergy spectrum for an inhomogeneous magnetic field in a stripe geometry.
More specifically, we assume a magnetic flux $\phi=-1/3$ inside a central stripe oriented along the $y$-axis, and $\phi=1/3$ outside it, with periodic boundary conditions along the horizontal direction, as schematically illustrated in the inset of Fig.~\ref{fig:bulk_and_edge_spectrum}.
The stripe geometry allows us to preserve translational invariance along $y$-direction, and to study the dispersion relation as a function of quasimomentum $k_y$.

The inhomogeneous magnetic field is realized by generalizing the definition of the magnetic field operator in Eq.~(\ref{eq:spin-dependent_phase_shift}), so as to allow position-dependent values of the magnetic field, $B(x,y)$.
In particular, to realize the stripe geometry, it is sufficient for the operator \encapsulateMath{$\hat{F}$} to imprint a spin-dependent linear phase gradient of opposite slope for the inner and outer regions, so that in these two regions a homogeneous flux of opposite sign is created.
The details how $B(x,y)$ is defined at the boundary are unimportant, since the topological properties we are interested in are robust, and do not depend on the detailed shape of the boundaries.
The only requirement is that the stripes are chosen sufficiently wide in order to have two well-defined bulk regions, the central stripe and the surrounding one, separated by two straight edges extending along the $y$-direction.

Relying on translational invariance along the $y$-axis, we show in Fig.~\ref{fig:bulk_and_edge_spectrum} the quasienergy spectrum plotted as a function of quasimomentum $k_y$.
The continuum of states, indicated by filled regions, coincides with the bulk quasienergy bands shown in Fig.~\figref{fig:quasienergy_spectrum_bulk}{b}.
In addition, in the quasienergy gaps, continuous branches of the dispersion relation can be clearly identified, which represent the \TP edge modes of the magnetic quantum walk for the stripe geometry considered here.
This example allows us to demonstrate how the bulk-boundary correspondence applies to magnetic quantum walks, based on the RLBL gap invariants, $R$, which are defined in Sec.~\ref{sec:top_invariants}.
The inside and outside regions of the stripe have related RLBL invariants, since these two regions only differ in the sign of the magnetic flux:
In fact, under flux inversion, the position of the quasienergy band gaps remains the same, but the gap invariants change sign.
A proof of that is provided in Appendix~\ref{app:flux_inversion}.
Hence, considering the values of $R$ provided in Fig.~\figref{fig:quasienergy_spectrum_bulk}{b} for $\phi=1/3$, the bulk-boundary correspondence is easily verified: for each gap, the net number of edge modes associated with a given edge, with the upward- and downward-propagating modes counted with opposite sign, matches exactly the difference between the RLBL invariants of the two bulk regions.
We also note that all quasienergy gaps in the example shown in the figure host at least one topologically protected edge mode, demonstrating the presence of anomalous edge modes with their quasienergy winding in the Floquet zone \cite{Rudner2013,Titum:2016}.

To illustrate the remarkable robustness of edge modes, we investigate magnetic quantum walks for a 
magnetic-field landscape, which includes irregular boundaries between different magnetic domains.
We consider a magnetic field with a constant flux $\phi=1/3$ everywhere, except for inside an ``island'' shaped as a quarter of a circle, where the flux is inverted, $\phi=-1/3$.
We consider in particular the evolution of a
walker starting from a single site close to the boundary of the island, since this allows us to excite the 
edge modes and study their transport properties \cite{Groh2016}.
In view of future experiments, we 
simulate the walker's evolution assuming realistic experimental conditions, taking into account the finite optical resolution of the imaging system used to create the artificial vector potential, as explained later in Sec.~\ref{sec:hommagfield}.
Moreover, we focus our attention on the spatial probability distribution of the walker, an observable that is readily accessible experimentally.

The simulated probability distribution is shown in Fig.~\ref{fig:evolution_along_quarter_circular_boundary} for an increasing number of steps.
During the time evolution, the walker's wavefunction stays mostly in the vicinity of the boundary, with only a small fraction of it expanding into the bulk regions.
Focussing on the edges, the wavefunction splits into two wave packets moving clockwise and counterclockwise along the boundary. 
The reason is that the initial state of the walker has a nonvanishing overlap with the edge modes of all quasienergy gaps, and these modes have different propagation direction depending on the sign of the respective gap invariant.
Once the wave packets are clearly separated, 
we observe that they travel across the sharp corners without back-scattering nor propagating into the bulk (see also the animation in the Supplemental Material \cite{VideoBoundary}),    
thus showing the remarkable robustness of unidirectional transport through \TP edge modes.

\section{Experimental realization} \label{sec:generating_artifial_magnetic_field}
We suggest using ultracold Cs atoms in a state-dependent optical lattice for the experimental realization of discrete-time quantum walks on a square lattice.
Currently, there is an ongoing work in the Bonn quantum-walk laboratory to extend the one-dimensional quantum-walk scheme \cite{Karski2009,Robens:2015} from one to two dimensions.
Recently, arbitrary state-dependent shift operations, \encapsulateMath{$\hat{S}_x$} and \encapsulateMath{$\hat{S}_y$}, have been demonstrated, thus realizing the two main operations required for 2D discrete-time quantum walks.
These experimental results will be reported elsewhere.

In Sec.~\ref{sec:two_dim_quantum_walks}, we shall give a brief account of the experimental setup realizing 2D discrete-time quantum walks, referring the reader to Ref.~\cite{Groh2016} for more details on the 2D state-dependent transport scheme. The remaining two sections, \ref{sec:artificialgaugefieldimp} and \ref{sec:hommagfield}, are devoted to the implementation of the magnetic-field operator $\hat{F}$, as defined in Eq.~\eqref{eq:spin-dependent_phase_shift}.

\subsection{Two-dimensional quantum-walk setup} \label{sec:two_dim_quantum_walks}

We use two hyperfine ground states of cesium atoms to represent the two spin states of the walker, $\ket{\uparrow} = \ket{F = 4, m_{F} = 3}$ and $\ket{\downarrow} = \ket{F = 3, m_{F} = 3}$.
The atoms are trapped, depending on the spin state, in two independent square optical lattices, with lattice constant $a=\lambda_{L}/\sqrt{2}$ \cite{Groh2016}, originating from right- and left-handed circularly polarized light.
In fact, at the wavelength of $\lambda_{L}\sim \SI{870}{\nano\meter}$, atoms in $\ket{\uparrow}$ and $\ket{\downarrow}$ state experience only the attractive optical dipole potential produced by right- and left-handed circularly polarized light, respectively.
These two lattices are spatially overlapped on the same plane and individually controlled by a high precision polarization synthesizer \cite{Robens2018}.
The optical lattice depth $V_0$ is chosen sufficiently deep, typically at around a thousand recoil energies $E_R$, so as to suppress site-to-site tunneling and, concurrently, to allow fast shift operations.
Initially, a number of atoms can be individually arranged \cite{Robens2017} in well-defined lattice sites, where they are cooled into the lowest motional state (i.e., lowest energy band) by means of sideband cooling techniques.

The shift operators, $\hat{S}_x$ and $\hat{S}_y$, are implemented by rotating by $\SI{180}{\degree}$ the linear polarization of either one of the two polarization-synthesized lattice beams forming the 2D spin-dependent optical lattice \cite{Groh2016}.
As a result of that, the two optical lattices, which selectively trap atoms in either $\ket{\uparrow}$ or $\ket{\downarrow}$ state, are shifted by half lattice site in opposite directions along the $x$- or $y$-axis, respectively.
Importantly, the shift operators must be performed fast, so as to outrun decoherence produced by fluctuating magnetic field (limiting coherence time to $\lesssim\SI{10}{\milli\second}$ \cite{Ruster:2016}) and by spontaneous scattering of lattice photons (limiting coherence time to $\lesssim\SI{100}{\milli\second}$) \cite{Alberti:2014}.
Using quantum optimal control \cite{Caneva:2009}, or shortcuts to adiabaticity \cite{Torrontegui:2013} or, in general, a non-adiabatic control scheme, the shift operators can be realized without creating motional excitations \cite{Thau:2019} on a short time scale, of the harmonic trap period $\tau_\text{HO} = \sqrt{m \lambda_L^2 /V_0}$, which for sufficiently deep lattices can be of the order of \SI{10}{\micro\second}.
This transport time is about two orders of magnitude shorter than the time required for site-to-site tunneling ($\gtrsim \hbar/E_R$) in a shallow optical lattice.

To implement the global coin operator, $\hat{C}$, we use microwave $\pi/2$ pulses that are resonant with the hyperfine energy splitting between the two spin states ($\Delta_\text{HF}=\SI{9.2}{\giga\hertz}$), as done in previous experiments \cite{Karski2009,Robens:2015}.
The 2D quantum-walk apparatus achieves a Rabi frequency of about \SI{200}{\kilo\hertz}, allowing the Hadamard-like coin operator $\hat{C}$ to be realized in  $\sim\SI{1}{\micro\second}$.

\subsection{Implementing artificial gauge fields}\label{sec:artificialgaugefieldimp}

A natural way to implement the magnetic-field operator $\hat{F}$, as defined in Eq.~\eqref{eq:spin-dependent_phase_shift}, consists of applying onto the atoms a spin-dependent linear potential gradient  
for a fixed duration, in such a way that
a spin-dependent
phase gradient,
$\sigma_z\hspace{0.3pt} B \hspace{0.5pt} x$, is imprinted onto the wavefunction of the walker (with $x\in \mathbb{Z}$ denoting the lattice site coordinate in the $x$-direction).

To realize the spin-dependent 
potential,
one could simply use a (real) magnetic field gradient to induce a linear Zeeman energy shift.
In practice, however, this possibility must be excluded, since exorbitantly large magnetic field gradients, $>\SI{e3}{\gauss/\centi\meter}$, are needed in order to realize $\hat{F}$ in a time interval of $\sim\SI{10}{\micro\second}$, 
not to mention the difficulties involved with switching on and off the magnetic field gradient in such a short time.

We instead suggest to flash a light field for a fixed duration in order to realize the operator $\hat{F}$.
The potential induced by the light field must be able to discriminate between the two electronic spin states, $\ket{\uparrow}$ and $\ket{\downarrow}$.
For this purpose, there are two physical mechanisms in neutral atoms that can mediate an interaction between the electronic spin and a light field: (M1) the atomic spin-orbit coupling and (M2) the hyperfine interaction between the nuclear and the electronic spin. 
In the following, we shall see that both of them can be used to realize the operator $\hat{F}$.

\begin{figure}[t]
\begin{center}
\includegraphics[width=\columnwidth]{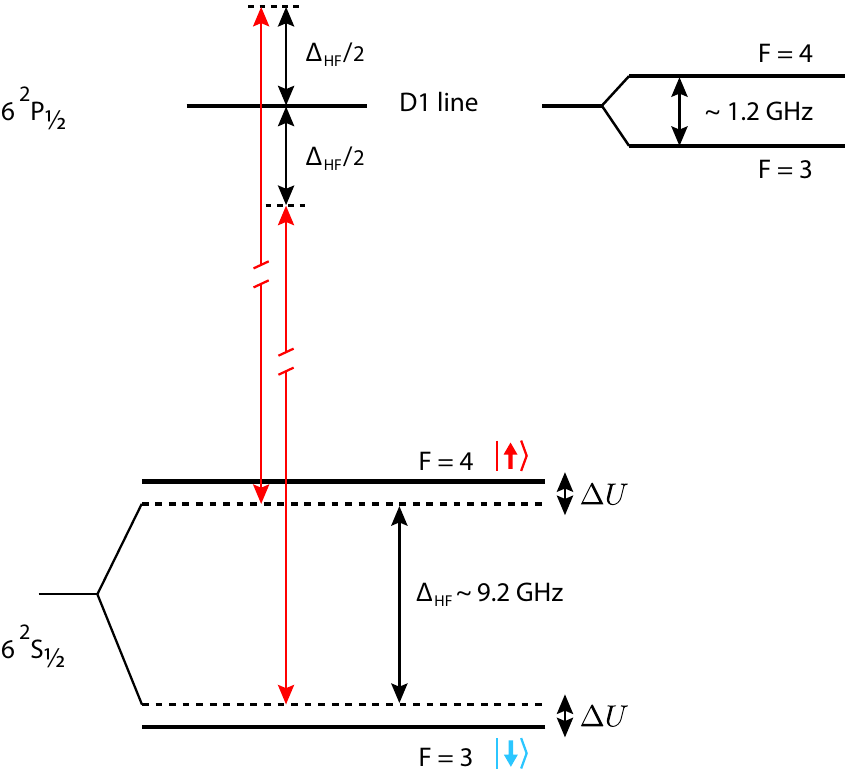}
\end{center}
\caption{Schematic representation of Cs energy levels corresponding to the
 D1 transition, including the related hyperfine structure.
A laser beam is detuned by $\pm \Delta_\textrm{HF}/2$ from resonance with the $\ket{\uparrow}$ and $\ket{\downarrow}$ state, respectively, so that the induced optical potential shifts the two states by an equal amount, $\Delta U$, but in opposite directions.
Short light pulses, thus, imprint a purely differential phase shift onto the two states.
A similar scheme exists for the D2 transition.
}
\label{fig:Cs atom fine and hyperfine structure}
\end{figure}

In the approach based on 
the interaction mediated by the spin-orbit coupling (M1),
the atoms are illuminated with a circularly polarized light field tuned at the magic wavelength  $\lambda_M= \SI{880}{\nano\meter}$, corresponding to a laser frequency lying in between the D1 and D2 line of Cs atoms.
At this special wavelength, the light field yields a purely differential light shift of the two internal states, meaning that $\ket{\uparrow}$ and $\ket{\downarrow}$ states are shifted in energy by the exact same amount, but in opposite directions.
However, for this to work, the quantization axis must be slightly tilted (e.g., by \SI{15}{\degree}) off the optical lattice plane, in such a way that the polarization of the light field has a nonzero circular component with respect to the direction of the quantization axis.

In the alternative approach based on 
the hyperfine interaction between the nuclear and the electronic spin (M2), 
one illuminates the atoms with a linearly polarized light field with its wavelength, $\lambda_M$, corresponding to a laser frequency tuned in the proximity of either the D1 or D2 line of Cs atoms.
By tuning $\lambda_M$ precisely in between the hyperfine structure, this second approach, too, allows one to realize a purely differential light shift of the two spin states, as shown in Fig.~\ref{fig:Cs atom fine and hyperfine structure}.
The advantage of this second approach is that it also applies to the case in which the quantization axis is in the same plane of the optical lattice \cite{Groh2016}.

In both approaches, the magnitude of the energy shift at site $(x,y)$ is determined by the local intensity of the light field.
Flashing such a light field for a fixed duration imprints onto the two spin states a differential phase shift, which is proportional to the product of the light field intensity and its pulse duration.
Therefore, to realize a homogeneous artificial magnetic field, it is in principle sufficient to project onto the atoms a light field with the intensity increasing linearly in the $x$-direction.

However, the same light field used to implement the magnetic-field operator can have undesired side effects, leading to decoherence of the quantum walker.
In what follows, we discuss the two main sources of decoherence involved in this process, motional excitations and photon scattering, and present solutions how to avoid them.

\begin{figure}[t]
\includegraphics[width=\columnwidth]{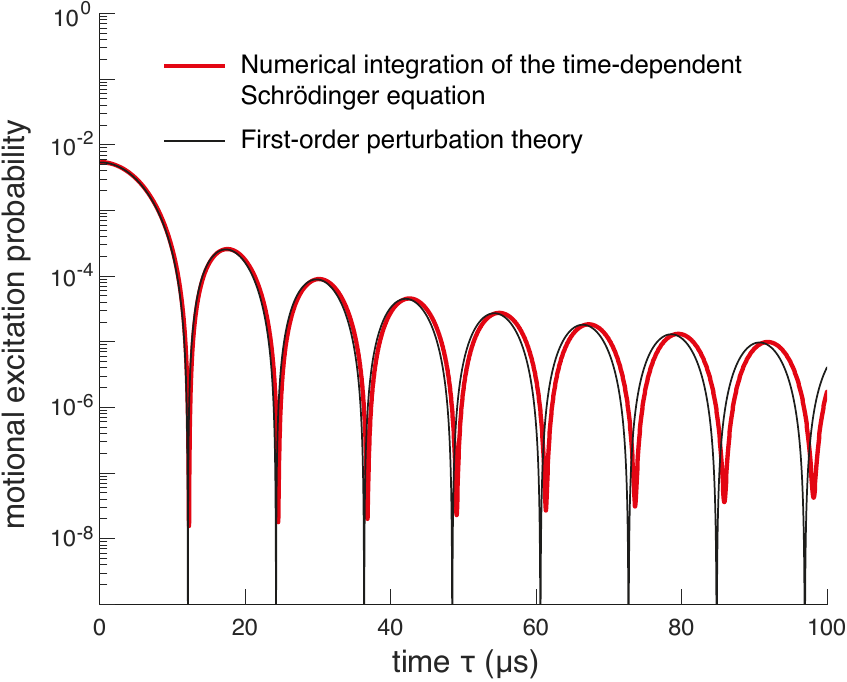}
\caption{Probability to create a motional excitation to a higher energy band of the optical lattice, after stroboscopically applying for a time $\tau$ the spin-dependent linear potential gradient that realizes the magnetic-field operator $\hat{F}$.
We assume here a lattice depth $V_0\sim 850\,E_R$, a field flux $\phi=1/3$, and that the atom is initially prepared in the motional ground state. The two curves show the probability obtained through numerical integration of the time-dependent Schrödinger equation (thick red curve), and the probability to excite an atom to the first motional state obtained using Eq.~(\ref{eq:probexcitations}) (thin black curve).
The small discrepancies between the two curves are ascribed to the anharmonicity of the optical lattice potential.
}
\label{fig:MotionalExcitations}
\end{figure}

	\paragraph{Decoherence by motional excitations.}
	Switching the linear potential gradient stroboscopically on and off can in principle excite the atoms to higher vibrational states (i.e., higher lattice bands), thus resulting in fast decoherence of the quantum walk.
	In fact, assuming for simplicity that the spin-dependent linear potential gradient is suddenly switched on for a finite time, $\tau$, during this time the atomic wave packet is subject to a perturbative potential, $\hat{H}_\phi$, which in the limit of deep lattices ($V_0\gg E_R$) reads as
	\begin{equation}
		\label{eq:hamilt_force}
		 \hat{H}_\phi \approx \frac{2^{1/4}\,\phi}{(E_R/V_0)^{1/4}}\,\frac{\hbar}{\tau}\,(\hat{b}^\dagger+\hat{b})\,\hat{\sigma}_z, 
	\end{equation}
	where $\hat{b}^\dagger$ and $\hat{b}$ are the operators creating and annihilating motional excitations (see Appendix~\ref{app:perturbative_force}).
		Using first-order time-dependent perturbation theory, the probability to excite an atom from motional ground state to the first excited state can be estimated as:	
	\begin{equation}
		\label{eq:probexcitations}
	 p_\text{ex}\approx\frac{\sqrt{2}\,\phi^2  \sinc^2(\pi \tau / \tau_\text{HO})}{\sqrt{V_0/E_R}}.
	\end{equation}
	Figure~\ref{fig:MotionalExcitations} shows that for deep lattices, this probability is very small, $<\num{e-2}$.

	Moreover, motional excitations can be further suppressed by a significant amount by tuning the duration, $\tau$, at a multiple of the harmonic trap period, $\tau_\text{HO}$.
	In this case, however, to estimate the number of residual motional excitations, one must go beyond the harmonic approximation assumed to derive Eq.~(\ref{eq:probexcitations}).
	For this purpose, we have computed the number of motional excitations by integrating numerically the time-dependent Schrödinger equation using the split-step finite difference propagation method \cite{Feit:1982}.
	Our results show that the probability of motional excitations can be reduced to $<\num{e-7}$.
	It is worth remarking that quantum control theory \cite{Caneva:2009} and shortcuts to adiabaticity \cite{Torrontegui:2013} can be employed to speed up the phase imprint process while avoiding motional excitations, and to make the dynamics less sensitive to small drifts of the experimental parameters.

\paragraph{Decoherence by photon scattering.}
		Projecting onto the atoms a linear intensity gradient, as previously suggested for realizing $\hat{F}$, has the negative effect of exposing the atoms, depending on their position in the lattice, to a high-intensity light field.
	Since the probability of scattering a photon by an atom is proportional to the intensity, a high-intensity light field would lead to a high photon scattering rate and, thus, to severe spatial decoherence \cite{Alberti:2014} of the quantum walk.
					In the following section, we show how high photon scattering rates can be entirely avoided by taking advantage of the Floquet dynamics.

\subsection{Floquet phase imprinting to avoid photon scattering}\label{sec:hommagfield}
\begin{figure}[t]
 \centering
  \includegraphics[width=\columnwidth]{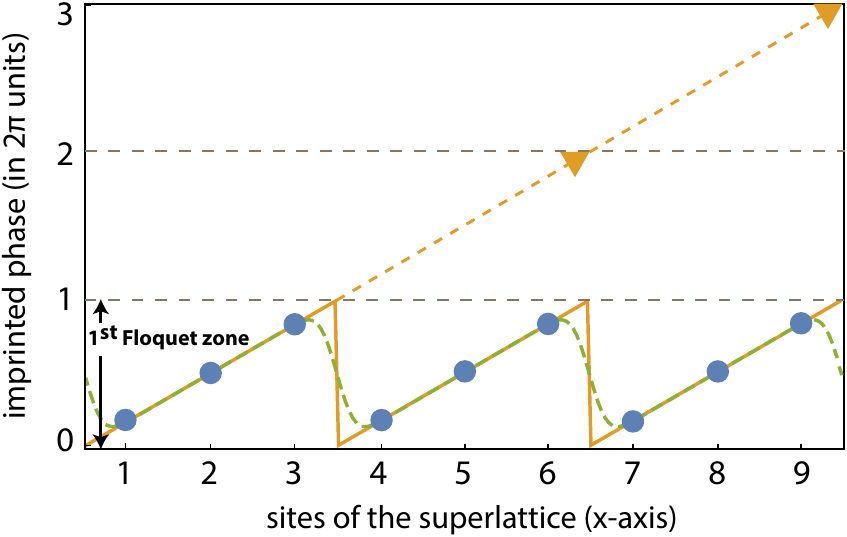}
  \caption{Floquet phase imprint realizing the magnetic-field operator $\hat{F}$.
Only the phase for one spin component is shown in the figure, the other one being equal in magnitude, but opposite in sign.
The graph shows the representative situation of a magnetic flux $\phi=1/3$, which is realized in the Landau gauge $\vect{A} = (0,Bx,0)$.
Instead of imprinting a linear phase gradient (dashed orange line), a sawtooth-shaped phase pattern (dashed green line) lying within the first Floquet zone, $[0,2\pi]$, is imprinted onto the atoms (note the different definition here of the Floquet zone with respect to Sec.~\ref{subsection:bulk_spectrum}).
The modulated phase pattern is produced by a sawtooth intensity profile, which is imaged onto the atoms through a high-numerical-aperture objective lens $\text{NA}=0.92$ \cite{Robens2017a}.
Due to the $2\pi$-periodicity of the imprinted phases, the two phase patterns are effectively identical in the proximity of the atoms (blue dots).
The folded phase pattern is also shown in the idealized case of no optical diffraction (solid orange line).
 }
\label{fig:SawtoothIntensityPattern}
\end{figure}

To avoid exposing the atoms to a high-intensity light field, we propose to use a sawtooth-shaped intensity pattern to illuminate the atoms, instead of the linear intensity gradient discussed previously.
The scheme proposed here allows one to realize the same magnetic-field operator, $\hat{F}$, while, at the same time, it avoids the problem of decoherence \cite{Alberti:2014} by photon scattering.
The basic intuition underlying the suggested scheme is the fact that the artificial-magnetic-field landscape realized by the operator $\hat{F}$ is determined by the phases imprinted onto the quantum walker's wavefunction.
Therefore, by folding the light field in such a way as to avoid phases exceeding $2\pi$, 
 we can realize the same operator $\hat{F}$, while avoiding high-intensity light fields.
 Figure~\ref{fig:SawtoothIntensityPattern} provides an illustration of the proposed scheme for the case of $\phi=1/3$.

Experimentally, a sawtooth modulation of the intensity can be realized using a spatial light modulator \cite{Zupancic:2016} combined with a high-resolution imaging system \cite{Robens2017a}.
The intensity of the light field depends, in particular, on which of the two approaches, M1 or M2 (see Sec.~\ref{sec:artificialgaugefieldimp}), is used to produce artificial magnetic fields.
For example, to imprint in \SI{10}{\micro\second} a spin-dependent phase shift of the order of $2\pi$ (larger phase shifts are not needed) over an area of $\mathcal{A}=100\times 100$ lattice sites, we estimate \cite{Grimm2000} that the atoms must be illuminated with a light field of approximately $\SI{40}{\milli\watt}/\mathcal{A}$ and $\SI{20}{\micro\watt}/\mathcal{A}$ for the M1 and M2 driving schemes, respectively.
Using the Kramers-Heisenberg formula, we also estimate the probability of an atom to scatter a photon of the light field.
We find that during each step of the magnetic quantum walk this probability is approximately $\num{e-5}$ in the approach M1, and $\num{e-3}$
in the approach M2.
This means that, on average, many steps of the quantum walk can be carried out before a photon is scattered: several tens of thousands in case of M1 and a few hundreds in case of M2.
Note that the disparity between these two values is due to the different strength of the two mechanisms, (M1) and (M2), which mediates the interaction between the electron spin and the light field.
It is also worth remarking that the scattering probability does not depend on the laser intensity \cite{scattering_prob}
nor on any other tunable parameters, but is simply determined by the atomic properties of Cs atoms.

For an experimental realization of the Floquet phase-folding scheme, one should take into account the finite optical resolution of the imaging system, whose effect is to smoothen the intensity pattern imaged onto the atoms. With reference to Fig.~\ref{fig:SawtoothIntensityPattern}, the imaged sawtooth intensity profile exhibits soft falling edges, which extend over a length scale comparable to that of the point spread function of the imaging system.
For a diffraction-limited imaging system, the size of the edges corresponds to the Abbe radius, $\lambda_M/(2\text{NA})$, which for a high numerical aperture, $\text{NA}\lesssim1$, can be slightly smaller than the separation between two adjacent sites, $\lambda_L/\sqrt{2}$.
Figure~\ref{fig:SawtoothIntensityPattern} shows indeed that using a high-numerical-aperture imaging system it is possible to accurately simulate a magnetic flux of $\phi=1/3$ in spite of the limited optical resolution.
Also, it is worth emphasizing that the numerical studies presented in Fig.~\ref{fig:bulk_and_edge_spectrum} and Fig.~\ref{fig:evolution_along_quarter_circular_boundary} are computed using the Floquet phase-folding scheme introduced in this section, assuming a numerical aperture $\text{NA}=0.92$ \cite{Robens2017a}.

\begin{figure}[b]
	\centering
	\includegraphics[width=\columnwidth]{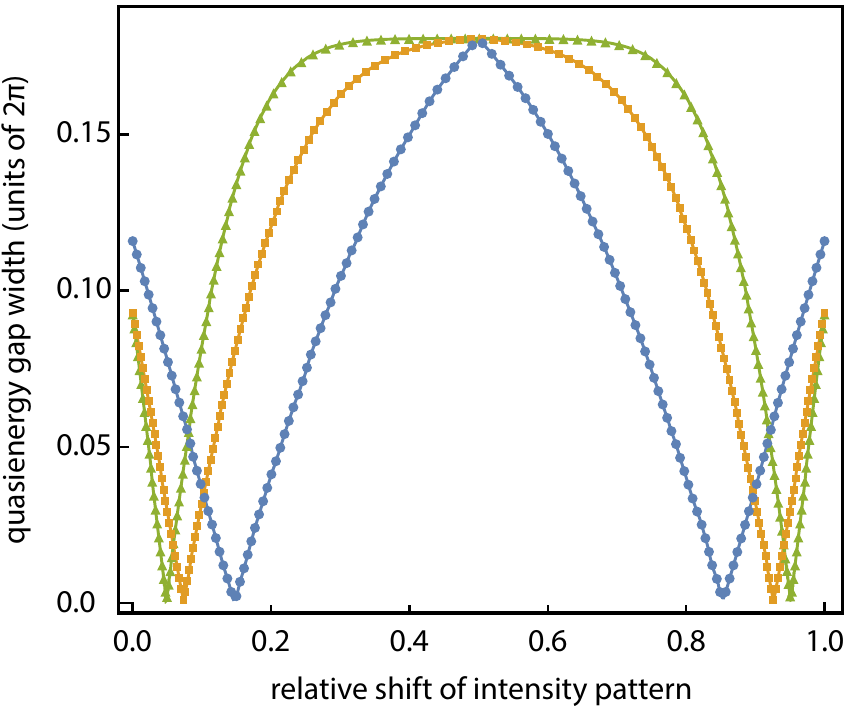}
	\caption{Quasienergy gap width of the bulk spectrum as a function of the relative shift between the sawtooth intensity profile and the optical lattice.
All quasienergy gaps turn out to be identical.
The graph is shown for the representative case of a magnetic flux $\phi=1/3$.
The three curves refer to the case of a simple lattice (\textcolor[RGB]{94,129,181}{$\bullet$}), a superlattice with a two-fold longer lattice constant (\textcolor[RGB]{255,156,36}{\mbox{\tiny $ \blacksquare $}}), and a superlattice with a four-fold longer lattice constant (\textcolor[RGB]{143,176,50}{$\blacktriangle$}).
The relative shift is expressed in units of the rescaled lattice constant.
The quasienergy gap width is maximum at a relative shift of $0.5$, when the sawtooth intensity profile is aligned with respect to the lattice sites as shown in the illustration in Fig.~\ref{fig:SawtoothIntensityPattern}.  
} \label{fig:gapwidthVsshift}
\end{figure}

The soft falling edges of the sawtooth intensity pattern represent a possible source of error for the magnetic-field operator.
If a lattice site happens to be centered on such an edge, and the walker's wavefunction extends on that site, the phase imprinted on that part of the wavefunction will differ from the one of a homogeneous magnetic field. 
This inhomogeneity in the magnetic field can have severe consequences on the band structure of the magnetic quantum walk, 
possibly leading to a closing of the bulk quasienergy gaps, as shown in Fig.~\ref{fig:gapwidthVsshift}.
To avoid such defects in the imprinted phases, it is required 
that the sawtooth intensity profile is carefully aligned with respect to the optical lattice, as shown in the illustration in Fig.~\ref{fig:SawtoothIntensityPattern}.

A way to make the proposed scheme more robust against alignment errors is to carry out the magnetic quantum walks on a superlattice.
This can be realized by replacing in the quantum-walk protocol the shift operator \encapsulateMath{$\hat{S}_x$} by the ``super shift'' operator \encapsulateMath{$(\hat{S}_x)^m$}, consisting of a spin-dependent shift of atoms in the $x$-direction by a multiple $m$ of lattice sites.
Spin-dependent shift operations by a multiple of lattice sites can be readily implemented with polarization-synthesized optical lattices \cite{Robens2017} by exploiting the reset-free control of linear polarization \cite{Robens2018}, which enables a continuous rotation of the polarization by $m$ times $\SI{180}{\degree}$ (compare with Sec.~\ref{sec:two_dim_quantum_walks}).
These supershift operators, applied to quantum-walk protocol defined in Eq.~(\ref{eq:main_1}), allow simulating magnetic quantum walks on a superlattice, whose lattice constant is effectively augmented by a factor $m$ compared to the original lattice constant, $a$.
A longer lattice constant has the advantage to yield an effective $m$-fold increase of the spatial resolution with which arbitrary magnetic-field landscapes are engineered.
Importantly, the longer lattice constant 
only requires a slight increase in 
the duration of the shift operators, which 
is proportional to 
$\sqrt{m}$ \cite{Thau:2019}.
In contrast, in conventional optical lattice systems relying on site-to-site tunneling, a similar $m$-fold increase of the lattice constant would result in a significant reduction of the tunneling rate, which scales exponentially with $m$.
We expect for an effectively larger lattice constant a higher robustness of the experimental scheme against experimental misalignments.
To quantify such a robustness enhancement, we analyze the size of the quasienergy gaps as a function of the relative shift of the sawtooth intensity profile with respect to the optical lattice.
The numerical results, which are displayed in Fig.~\ref{fig:gapwidthVsshift}, reveal that a superlattice with $m=4$ makes the topological structure largely insensitive to alignment errors.
This confirms our intuition, showing that an $m$-fold increase of the lattice constant in the $x$-direction is equivalent to an enhancement of the optical resolution of the imaging system by the same factor.

Finally, it is worth emphasizing that the Floquet phase-folding scheme of Fig.~\ref{fig:SawtoothIntensityPattern} is not limited to the case of a homogeneous artificial magnetic field, as primarily discussed in this section.
It is straightforward to extend this scheme to situations where the sawtooth intensity profile has a different slope, or even a different orientation, in different regions of the lattice.
This allows one to create boundaries between different topological phases as those studied in Fig.~\ref{fig:evolution_along_quarter_circular_boundary}.

\section{Conclusions}

We have studied the magnetic \DTQW{} of a particle with two internal states, moving in discrete steps on a square lattice.
By calculating the Chern numbers of the bands, and complementing these with the RLBL invariants of the Floquet gaps, which are specific to periodically driven systems,
we could show that a magnetic quantum walk behaves like an anomalous Floquet--Chern insulator. 
Alongside the analysis of the bulk topological invariants, we have considered the scenario of inhomogeneous magnetic fields, where magnetic domains with different magnetic fluxes are connected through sharp edges, and studied the \TP edge modes formed along them.
This study has shown that the bulk-boundary correspondence applies to magnetic quantum walks, where the number of \TP edge modes in a given gap corresponds to the difference of RLBL gap invariants associated with the two magnetic domains interfacing at the boundary.
Moreover, by considering irregular boundaries between magnetic domains, we could show that the excitation of \TP edge modes enables the robust transport of the walker along the entire length of a magnetic domain, even in spite of sharp corners in its shape.

For the experimental realization of magnetic quantum walks, we have proposed a realistic scheme based on ultracold cesium atoms trapped in state-dependent optical lattices.
A remarkable aspect of the proposed scheme is that it allows us to generate artificial vector potentials with arbitrary landscapes.
This gives the flexibility to change the direction of the simulated magnetic fields and to create different topological domains with sharp spatial boundaries.
The Floquet nature of the system makes it possible to engineer any arbitrary gauge field with low intensities of the laser fields, thus significantly suppressing the probability of off-resonant photon scattering by the atoms.
This includes the possibility of creating uniform magnetic fields in the strong field regime of the Hofstadter butterfly spectrum, with no need for field rectification protocols, nor for any readjustment of the configuration of the laser beams in order to tune the field strength \cite{Aidelsburger:2017}.
We remark that state-dependent transport allows us to transport both spin states over several lattice sites making it possible to work with superlattice potentials, shortening significantly the evolution time with respect to schemes based on laser-assisted tunneling, which are limited by damping of tunneling rates.
In addition, the full control of micromotion in a magnetic quantum walk, where each operator constituting \encapsulateMath{$\hat{W}$} can be precisely realized in experiments, provides a new route to overcome the problem of heating in periodically driven systems \cite{Eckardt2017}.

An interesting avenue for future research consists in generalizing magnetic quantum walks to many-particle systems, where the atoms are allowed to interact with each other through contact potentials \cite{Ahlbrecht:2012}.
The quasienergy bands of the magnetic quantum walks are relatively flat, well separated by gaps, and characterized by nonvanishing Chern numbers. 
These constitute favorable conditions to realize strongly correlated states, which could lead to fractional Floquet Chern insulators \cite{Bergholtz:2013,Parameswaran:2013,Neupert:2015}.
Understanding the properties of such strongly correlated systems is one of the main goals in the field of quantum simulation \cite{Maciejko:2015}.
It remains, however, an outstanding theoretical and experimental challenge to find a way to ensure full population of a given Floquet band of the magnetic quantum walk. 

Finally, it is interesting to remark that the proposed scheme can be readily extended to simulate concurrently magnetic and electric fields.
Electromagnetic quantum walks can, in fact, be realized by slightly detuning the wavelength of the light field used to implement the operator $\hat{F}$ off its magic value, $\lambda_M$.
Thereby, the light field produces a combination of differential and common-mode phase shifts, which control the artificial magnetic and electric field, respectively.
In recent years, electric quantum walks been the subject of much research \cite{Wojcik:2004,Regensburger:12,Genske2013,Cedzich:2013,Xue:2014}, which could be extended to achieve new form of control of topological states of matter, especially through time-varying artificial electromagnetic fields \cite{Wang2018}.

\begin{acknowledgments}
A.A.\ acknowledges insightful discussions with M.\ Fleischhauer and H.\ Kroha.
We thank P.\ Arnault for early contributions to this work and numerous discussions.%
We also thank T.\ Groh's assistance for the estimate of the motional excitations while flashing the magnetic-field operator. 
We acknowledge financial support from the ERC grant DQSIM (Project Nr.\ 291401), and the collaborative research center \mbox{OSCAR} funded by the Deutsche Forschungsgemeinschaft (Project Nr.\ 277625399 -- TRR 185). M.S. also acknowledges support from the Deutscher Akademischer Austauschdienst.
J.K.A.\ acknowledges support from the National Research, Development and Innovation Fund of Hungary within the Quantum Technology National Excellence Program (Project Nr.\ 2017-1.2.1-NKP-2017-00001), and FK 124723.
\end{acknowledgments}

\appendix
\makeatletter
\renewcommand{\@hangfrom@section}[3]{#1\@if@empty {#2}{#3}{\MakeTextUppercase{#2}\@if@empty {#3}{}{:\ \MakeTextUppercase{#3}}}}
\makeatother

\textheight=1.01\textheight
\section{General two-dimensional vector potentials}\label{artificial_vector_potential}

We briefly summarize how a general 2D magnetic vector potential can be realized with a quantum walk \cite{Yalcinkaya2015,Marquez-Martin:2018,Cedzich:2019}. 
To create an artificial vector potential $\vect{A}$ with both its $x$- and $y$-components nonvanishing, two magnetic-field operators are employed in the time-step operator:
\begin{align}
	\label{eq:main_2}
\hat{W} = \hat{F}_y \, \hat{S}_y \, \hat{C} \, \hat{F}_x \, \hat{S}_x \, \hat{C},
\end{align}
where $\hat{F}_d$ is the magnetic-field operator, which is applied after a shift along the $d$-direction.
The two magnetic-field operators, which are defined as the line integral of the artificial vector potential,
\begin{multline}
	\label{eq:magneticoperator}
\hat{F}_d = \sum_{\vect{r}}\bigg[\exp\left( i \int_{\vect{r}-\vect{e}_d}^{\vect{r}} \mathrm{d}\vect{r'} \cdot \vect{A}(\vect{r'}) \right)\ket{\vect{r}}\bra{\vect{r}}\otimes\ket{\uparrow}\bra{\uparrow}+\\
 \exp\left( i \int_{\vect{r}+\vect{e}_d}^{\vect{r}} \mathrm{d}\vect{r'} \cdot \vect{A}(\vect{r'}) \right)\ket{\vect{r}}\bra{\vect{r}}\otimes\ket{\downarrow}\bra{\downarrow}\bigg],
\end{multline}
imprint onto the walker's wavefunction the so-called Peierls phases \cite{Luttinger:1951}.
In Eq.~(\ref{eq:magneticoperator}), $\vect{e}_d$ is the lattice unit vector along the $d$-direction, $\vect{\hat{r}}=(\hat{x},\hat{y},0)$ is the lattice position operator taking discrete values, and $\vect{r'}=(x,y,0)$ is the integration variable taking continuous values in the lattice plane.

For the vector potential used in the main text, $\vect{A}(\vect{r})=(0,Bx,0)$, one finds that \encapsulateMath{$\hat{F}_x$} is equal to the identity operator, whereas \encapsulateMath{$\hat{F}_y$} is equal to \encapsulateMath{$\hat{F}$}, as defined in Eq.~(\ref{eq:spin-dependent_phase_shift}).
Hence, for this vector potential, the time-step operator in Eq.~(\ref{eq:main_2}) coincides with the operator defined in  Eq.~(\ref{eq:main_1}).

\section{Sublattice symmetries}
\label{app:sublattice_symmetries}

At each time step, the magnetic quantum walk changes the parity of both $x$- and $y$-coordinate. 
This results in two types of sublattice symmetries, which are discussed below. 
We caution that \textemdash unlike in a system described by a constant Hamiltonian \textemdash both of these sublattice symmetries are different from the chiral symmetry, which is discussed below in Appendix \ref{app:chiral_symmetry}.

\subsection{Conserved sublattice}

The lattice can be partitioned into two sublattices: 
\begin{align}
\label{eq:sublattice_index}
\text{I} & : (x+y) \module 2=0;\\
\text{II} & : (x+y) \module 2=1.\nonumber
\end{align}
As a result of the symmetric form of the shift operator \encapsulateMath{$\hat{S}_d$}, the time-step operator \encapsulateMath{$\hat{W}$} only couples states belonging to the same sublattice. 
Thus, we have two independent magnetic quantum walks, one taking place on sublattice I, the other on sublattice II, meaning that the sublattice index in Eq.~(\ref{eq:sublattice_index}) is conserved. 

For a magnetic quantum walk with uniform magnetic flux, $\phi=p/q$, the two quantum walks on sublattice I and II are identical. 
This results in a two-fold degeneracy of the energy spectrum, which is also known as fermion doubling \cite{RotheBook}.

In more detail, the Brillouin zone associated with each individual sublattice is two times smaller and skewed with respect to the Brillouin zone of the whole lattice.
The reciprocal primitive vectors of the sublattice Brillouin zone are: $\vect{g}_1=(1,0)\,2\pi/q$ and $\vect{g}_2=(0,1)\,\pi$ for even $q$, and $\vect{g}_1=(1,0)\,2\pi/q$ and $\vect{g}_2=(-1/q,1)\,\pi$ for odd $q$.
While $\vect{g}_1$ coincides with one of the reciprocal primitive vectors of the whole lattice, $\vect{g}_2$ is different and shorter.
It thus follows that the spectrum of the time-step operator can be folded in the reduced Brillouin zone defined by $\vect{g}_1$ and $\vect{g}_2$, where for any given quasimomentum, all quasienergy states are two-fold degenerate.
Hence, we conclude that for each eigenstate $\ket{\psi(\vect{k})}$ of \encapsulateMath{$\hat{W}$} with quasimomentum $\vect{k}$ and quasienergy $E$, there exists another eigenstate of the time-step operator with the same quasienergy $E$ and quasimomentum $\vect{k}+\vect{g}_2$; compare with Fig.~\ref{fig:quasienergy_spectrum_bulk} and Fig.~\ref{fig:bulk_and_edge_spectrum}.

Note that the case of zero flux, $\phi=0$, is equivalent to assuming $q=1$, meaning that the zero-field spectrum remains unchanged under a shift of quasimomentum by $\vect{g}_2=(-\pi,\pi)$.

\subsection{Alternating sublattice}

Both sublattices I and II introduced above can be further partitioned into two sub-sublattices each, defined as follows:
\begin{align}
	\label{eq:new_sub_lattice}
  \text{Ia}\,:&\, x \text{ even},\;y \text{ even};\\
  \text{Ib}\,:&\, x \text{ odd},\; y \text{ odd};\nonumber\\
  \text{IIa}\,:&\, x \text{ odd},\; y \text{ even};\nonumber\\
  \text{IIb}\,:&\, x \text{ even},\; y \text{ odd}.\nonumber
\end{align}
To represent the sublattice structure defined in Eq.~(\ref{eq:new_sub_lattice}), we introduce the operator $\hat\chi$: 
\begin{align}
\label{eq:chi_op_definition}
\hat{\chi} &= \sum_{x,y}
(-1)^y \ket{x,y}\bra{x,y}\otimes \hat{\identity},
\end{align}
which takes the value $1$ on sublattices Ia and IIa, and the value $-1$ on sublattices Ib and IIb.

The time-step operator \encapsulateMath{$\hat{W}$} only couples states belonging to different sublattices, namely, (Ia) $\leftrightarrow$ (Ib), and (IIa) $\leftrightarrow$ (IIb). 
From this, it immediately follows, that
\begin{align}
\hat{\chi} \hat{W}\hat{\chi}^\dagger &= - \hat{W}.
\end{align}
This, in turn, means that the effective Hamiltonian is transformed as $\hat{\chi}^\dagger \hat{H}_\text{eff}\hat{\chi} =  \hat{H}_\text{eff} + \pi$, modulo the Floquet zone.

The operator \encapsulateMath{$\hat{\chi}$} in Eq.~(\ref{eq:chi_op_definition}) does nothing but imprint a phase $\pi$ on the lattice sites with odd $y$ coordinate.
This corresponds to a shift of the quasimomentum operator $\hat{k}_y$ by $\pi$, modulo the Brillouin zone.
Hence, for a magnetic quantum walk with translational invariance along the $y$-direction, each of its eigenstates $\ket{\psi(k_y)}$ with quasienergy $E$ and quasimomentum $k_y$ must have a sublattice-symmetric-partner eigenstate, \encapsulateMath{$\hat{\chi}\ket{\psi(k_y)}$}, which again is an eigenstate of the time-step operator with quasienergy $E+\pi$, and displaced quasimomentum $k_y+\pi$; compare with Fig.~\ref{fig:quasienergy_spectrum_bulk} and Fig.~\ref{fig:bulk_and_edge_spectrum}.

\section{Chiral symmetry}
\label{app:chiral_symmetry}

We show that magnetic quantum walk possesses chiral symmetry.
This symmetry is important since it stabilizes the Dirac points appearing in the spectrum of magnetic quantum walks (see Appendices~\ref{app:dirac_cones} and \ref{app:top_charges}).
In fact, unlike Weyl points in three-dimensional systems, Dirac points in two dimensions can exist only in the presence of some stabilizing symmetry \cite{Goerbig:2017}.

A Floquet system has chiral symmetry if there exists a local unitary operator $\hat{\Gamma}$ transforming the time-step operator $\hat{W}$ as follows:
\begin{equation}
	\label{eq:chiral_sym}
	  \hat{\Gamma} \hat{W} \hat{\Gamma}^\dagger = \hat{W}^\dagger. \end{equation}
Correspondingly, the effective Hamiltonian [cf.~Eq.~(\ref{eq:Heff})] transforms as:
\begin{equation}
		\label{eq:chiral_sym_hamilt}
	 \hat{\Gamma} \hat{H}_\text{eff} \hat{\Gamma}^\dagger = - \hat{H}_\text{eff},
\end{equation}
meaning that if $\ket{\psi}$ is an eigenstate of the effective Hamiltonian with quasienergy $E$, then \encapsulateMath{$\hat{\Gamma}\ket{\psi}$} is also an eigenstate of the same Hamiltonian with quasienergy $-E$.

The time-step operator $\hat{W}$, as defined in Eq.~(\ref{eq:main_1}), does not have chiral symmetry.
However,
a suitable cyclic permutation of the operators constituting \encapsulateMath{$\hat{W}$}, the resulting time-translated time-step operator has chiral symmetry.
Importantly, such a cyclic permutation represents a shift of the time frame defining the single step of the walker, and has no effect on the physical properties of the walker, such as the energy spectrum. 

For magnetic quantum walks, there are actually two time-translated time-step operators having chiral symmetry:
\begin{equation}
	\label{eq:time_frames}
	\hat{W}'   =  \hat{W}_\uparrow \hat{W}_\downarrow,\quad \hat{W}''  =  \hat{W}_\downarrow \hat{W}_\uparrow.
\end{equation}
These are the product of the two half-step operators defined as follows:
\begin{align}
		\label{eq:half_step_operator1}
	\hat{W}_\uparrow &= \hat{F}_{\uparrow} \hat{S}_{y,\uparrow} \hat{C} \hat{S}_{x,\uparrow},\\
		\label{eq:half_step_operator2}
	\hat{W}_\downarrow &=  \hat{S}_{x,\downarrow} \hat{C} \hat{S}_{y,\downarrow} \hat{F}_{\downarrow},
\end{align}
where \encapsulateMath{$\hat{S}_{d,s}$} is the shift operator displacing in the $d$-direction the walker with spin state $\ket{s}$,
\begin{align}
	\hat{S}_{d,\uparrow}= \begin{pmatrix} e^{-i \hat{k}_d} & 0 \\ 0 & 1	\end{pmatrix},\quad 	\hat{S}_{d,\downarrow}= \begin{pmatrix} 1 & 0 \\ 0 & e^{i \hat{k}_d}	\end{pmatrix},
	  	  	  	  	  	  	  	  	  	  	  	  \end{align}
and \encapsulateMath{$\hat{F}_{s}$} is the magnetic field operator acting on $\ket{s}$,
\begin{align}
	\hat{F}_{\uparrow}= \begin{pmatrix} e^{i B \hat{x}} & 0 \\ 0 & 1 \end{pmatrix}, \quad 	\hat{F}_{\downarrow}= \begin{pmatrix} 1 & 0 \\ 0 & e^{-i B \hat{x}} \end{pmatrix}.
\end{align}
Using these operators, the shift operator defined in Eq.~(\ref{eq:shiftoperator}) can be expressed as \encapsulateMath{$\hat{S}_d = \hat{S}_{d,\uparrow}\hat{S}_{d,\downarrow}$}, whereas the magnetic field operator defined in Eq.~(\ref{eq:spin-dependent_phase_shift}) as \encapsulateMath{$\hat{F} = \hat{F}_{\uparrow}\hat{F}_{\downarrow}$}.
Using the following relations, 
\begin{equation}
	\hat{S}_{d,\downarrow}=\hat{\sigma}_x \hat{S}_{d,\uparrow}^\dagger \hat{\sigma}_x,\quad \hat{F}_{\downarrow}=\hat{\sigma}_x \hat{F}_{\uparrow}^\dagger \hat{\sigma}_x, \quad\hat{C}=\hat{\sigma}_x \hat{C}^\dagger \hat{\sigma}_x,
\end{equation}
it is straightforward to show that  \encapsulateMath{$\hat{W}_\downarrow=	\hat{\sigma}_x \hat{W}_\uparrow^\dagger \hat{\sigma}_x $}.
Thus we can express the time-translated time-step operators in a form that is visibly chiral symmetric, $\hat{W}'=\hat{\sigma}_x \hat{W}_\downarrow^\dagger \hat{\sigma}_x \hat{W}_\downarrow $ and $\hat{W}''=\hat{\sigma}_x \hat{W}_\uparrow^\dagger \hat{\sigma}_x \hat{W}_\uparrow $, with the chiral operator defined by
\begin{equation}
	\label{eq:chiral_sym_ope}
	\hat{\Gamma}=\hat{\sigma}_x.
\end{equation}
One can easily verify, in fact, that Eq.~(\ref{eq:chiral_sym}) applies for both \encapsulateMath{$\hat{W}'$} and \encapsulateMath{$\hat{W}''$}.

\section{Dirac points}\label{sec:chiral-symmetry}
\label{app:dirac_cones}
The quasienergy spectrum of the effective Hamiltonian, \encapsulateMath{$\hat{H}_\text{eff}$}, has Dirac points, i.e., conical band-touching points carrying a topological charge.
These are visible in Fig.~\ref{fig:quasienergy_spectrum_bulk}.

In the case of no magnetic flux, $\phi=0$, there are four such band-touching points, at quasimomenta $\vect{K}$ and quasienergy $E$:
	\begin{align}
	  \vect{K}_{\pm,\mp} &= \Big(\pm \frac{\pi}{2}, \mp \frac{\pi}{2} \Big), \quad E =0;  \\
	  \vect{K}_{\pm,\pm} &= \Big(\pm \frac{\pi}{2}, \pm \frac{\pi}{2} \Big),  \quad E =\pi. \label{eq:Dirac-Points-2DQW}
	\end{align}

Expanding the effective Hamiltonian around these points to the first order in the displaced quasimomentum operator, $\vect{\delta \hat{k}} = \vect{\hat{k}} - \vect{K}$, we obtain:
\begin{align}
	\vect{K}_{\pm,\mp}: \quad\hat{H}'_\text{eff}&\approx \mp \delta \hat{k}_x \hat{\sigma}_y + \delta \hat{k}_y \hat{\sigma}_z, \\ 	\vect{K}_{\pm,\pm}: \quad \hat{H}'_\text{eff}&\approx \pm \delta \hat{k}_x \hat{\sigma}_y + \delta \hat{k}_y \hat{\sigma}_z - \pi, \end{align}
which exhibits a Rashba-like spin-orbit coupling. 
Because of chiral symmetry, the eigenspinor, $(0,\mp \delta k_x, \delta k_y)$, of the effective Hamiltonian for $\vect{K}_{\pm,\mp}$ is confined to a vertical plane of the Bloch sphere, in which 
it winds once in a clock- or anticlockwise direction, depending on which of the two Dirac points, as the quasimomentum $(\delta k_x,\delta k_y)$ is varied along a closed contour containing the Dirac point. Similar considerations apply to the other two Dirac points $\vect{K}_{\pm,\pm}$.
The quasienergy dispersion in the vicinity of one of these Dirac points has a conical shape, $E(\delta k_x,\delta k_y) = E(k_x=0,k_y=0) \pm (\delta k_x^2+\delta k_y^2)^{1/2}$, indicating that the walker moves with a constant velocity modulus, which is equal to $a/T = 1$ (in the dimensionless units used in this work).

For completeness, we also report the effective Hamiltonian corresponding to the other chiral symmetric time frame, specified in the basis where the chiral operator \encapsulateMath{$\hat{\Gamma}$} is diagonal:
\begin{align}
	\vect{K}_{\pm,\mp}: \quad\hat{H}''_\text{eff}&\approx \pm \delta \hat{k}_y \hat{\sigma}_y+\delta \hat{k}_x \hat{\sigma}_z , \\ 	\vect{K}_{\pm,\pm}: \quad \hat{H}''_\text{eff}&\approx  \pm \delta \hat{k}_y \hat{\sigma}_y +  \delta \hat{k}_x \hat{\sigma}_z - \pi. \end{align}

In the case of a nonzero magnetic flux, we also find Dirac points.
For graphene-like lattices, Rhim and Park~\cite{rhim2012self} proved that Dirac points exist for any arbitrary magnetic flux.
We speculate that a similar result also applies to our situation.
This intuition is supported by the empirical observation that Dirac points carrying a nonvanishing topological charge exist for various choices of the magnetic flux $\phi=p/q$, varying $p$ and $q$ up to about 20, provided that $q$ is an odd number.
For a magnetic flux with even $q$, Dirac points merge, and their topological charges annihilate \cite{Goerbig:2017}.
However, the two bands remain touching in spite of the merge.
Using the effective Hamiltonians (restricted to the two touching bands), we have empirically observed that Dirac points at $E=0$ are associated with a topological charge (as defiend in Appendix~\ref{app:top_charges}) $\nu_0=1$ when $K_x>0$, and with $\nu_0=-1$ when $K_x<0$;
similarly, Dirac points at $E=\pi$ are associated with $\nu_\pi=-1$ when $K_x>0$, and with $\nu_\pi=1$ when $K_x<0$.

\section{Topological charges of Dirac points}
\label{app:top_charges}

We show that each Dirac point is associated with a winding number. In fact,
for any chiral symmetric Hamiltonian, such as the effective Hamiltonian of magnetic quantum walks (see Appendix~\ref{app:chiral_symmetry}), a winding number $\nu$ can be assigned to any closed loop $\mathcal{C}$ in the Brillouin zone, as long as the Hamiltonian is gapped along the entire loop.
This in particular holds for infinitesimally small loops surrounding a Dirac point.
The winding number can be defined considering the off-diagonal part $\hat{h}(\mathbf{k})$ of $\hat{H}_\text{eff}(\mathbf{k})$, with the latter denoting the effective Hamiltonian reduced to quasimomentum $\vect{k}$ and represented in the basis where the chiral symmetry operator $\hat{\Gamma}$ is diagonal, 
\begin{align}
	\label{eq:gamma}
	\hat{H}_\text{eff}(\mathbf{k}) = 
    \begin{pmatrix} 0 & \hat{h}(\mathbf{k}) \\ \hat{h}(\mathbf{k})^\dagger & 0 \end{pmatrix}; \quad \hat{\Gamma} &= \begin{pmatrix} \hat{\identity} & 0 \\ 0 & -\hat{\identity}  \end{pmatrix}.
\end{align}
The winding number then reads as
\begin{equation}
	\label{eq:winding_int}
	  \nu[\hat{h}] = \frac{1}{2\pi i} \oint_\mathcal{C} d\vect{k} \frac{d}{d\vect{k}} \log
	  \det \hat{h}(\vect{k}),
\end{equation}
with the line integral evaluated  along the loop $\mathcal{C}$ in the \BZ{}. 
This winding number is an integer, and is invariant under continuous deformations of the Hamiltonian $\hat{H}_\text{eff}$ or of the curve $\mathcal{C}$, since its value cannot change as long as chiral symmetry holds and the gap along $\mathcal{C}$ remains open.
This in particular applies to contours enclosing a Dirac point and excluding any other band-touching point, meaning that Dirac points are stable under small symmetry-preserving continuous deformations.
Such deformations may shift the position of the Dirac points in the \BZ{}, however, without lifting the degeneracy at the band-touching point, nor changing its quasienergy.
The degeneracy can only be lifted if two Dirac points with opposite topological charges meet and annihilate.

To obtain the topological charges of Dirac points in a magnetic quantum walk, one needs to calculate the winding number $\nu$ in both chiral symmetric time frames. 
This yields for each Dirac point two winding numbers, $\nu'$ from $\hat{W}'$, and $\nu''$ from $\hat{W}''$.
The topological charge $\nu_0$ of a Dirac point at quasienergy $E=0$ and 
$\nu_\pi$ of a Dirac point at quasienergy $E=\pi$ is then obtained by taking linear combinations of these winding numbers \cite{Asbth2013},
\begin{equation}
	\label{eq:top_charges}
	 \nu_0 = \frac{\nu'+\nu''}{2}, \quad \nu_\pi = \frac{\nu'-\nu''}{2}.
\end{equation}
For a Dirac point at $E=0$, we have $\nu_0 = \pm 1$ and $\nu_\pi=0$, while for a Dirac point at $E=\pi$, we have $\nu_\pi = \pm 1$ and $\nu_0=0$. 

Alternatively, one can use the half-step operator to compute the topological charges $\nu_0$ and $\nu_\pi$.
This second approach has the advantage that it does not require evaluating the effective Hamiltonian and, thus, is better suited for an analytical calculation of the winding number in Eq.~(\ref{eq:winding_int}), especially in weak-field regime, $q\gg1$, when the size of the magnetic unit cell is very large.
By representing the half-step operator in a block form in the basis where $\hat{\Gamma}$ is diagonal [see Eq.~(\ref{eq:gamma})],
\begin{equation}
 \hat{W}_\uparrow =
  \begin{pmatrix}
    \hat{a} & \hat{b} \\ \hat{c} & \hat{d}
  \end{pmatrix},
\end{equation}
one can show \cite{Asboth2014} that the topological charges defined in Eq.~(\ref{eq:winding_int}) correspond to $\nu_0 = \nu[b]$ and $\nu_\pi = \nu[d]$.

Moreover, it is worth noting that the winding number in Eq.~(\ref{eq:winding_int}) vanishes when evaluated on a contour enclosing the whole \BZ{}.
This has an intuitive explanation considering that each edge of the square magnetic \BZ{} contributes twice to the line integral, but with opposite sign, so that the overall integral vanishes.
The vanishing of the winding number for the whole \BZ{} allows us to conclude that the sum of the topological charges associated with the Dirac points at $E=0$ and $E=\pi$ respectively vanishes.
This also reflects the fact that the Dirac points occur in pairs (fermion doubling, see Appendix~\ref{app:sublattice_symmetries}).

\section{Topological invariants under flux inversion}
\label{app:flux_inversion}
We prove that under inversion of the magnetic flux, $\phi\rightarrow -\phi$, the Floquet topological invariants (as well as the Chern numbers) change sign.
The proof hinges on the behavior of the quantum walk under time reversal, and simultaneous inversion of the magnetic field over the whole lattice.
For the proof, we can generally assume an arbitrary, position-dependent magnetic field, i.e., a generic function $\phi(x,y)$, denoted below simply by $\phi$.

First, we note that the time-step operator transforms as
\begin{align}
\label{eq:particle_hole}
\hat{W}(\phi)^\ast&=\hat{W}(-\phi),
\end{align}
where the star symbol denotes the elementwise complex conjugation operation in the basis where the position operator, $\vect{\hat{r}}$, and $\hat{\sigma}_z$ are diagonal.
The transformation in Eq.~(\ref{eq:particle_hole}) holds in the original time frame of the magnetic quantum walk [see Eq.~(\ref{eq:main_1})], as well as in both chiral symmetric time frames [see Eq.~\ref{eq:time_frames})].
We note that Eq.~(\ref{eq:particle_hole}) specified for $\phi=0$ indicates that the quantum walk in a zero field has particle-hole symmetry, with the particle-hole operator represented by the complex conjugation.

Second, by combining the complex conjugation (i.e., the particle-hole operator) and chiral operator, we obtain an antiunitary operator, which defines \cite{Ryu:2010} the time-reversal operator of the quantum walk.
In any of the two chiral symmetric time frames (to be specific, we choose the one corresponding to \encapsulateMath{$\hat{W}'$}), the time-reversal operator transforms the time-step operator as:
\begin{align}
\label{eq:time_rev_transformation}
\hat{\Gamma} \hat{W}'(\phi)^\ast \hat{\Gamma}^\dagger
= \hat{W}'(-\phi)^\dagger,
\end{align}
showing that for $\phi=0$ the quantum walk has time-reversal symmetry, whereas for a non-vanishing magnetic field, the effective Hamiltonian is preserved under time-reversal operator, provided that the magnetic field is simultaneously reversed.

Third, it is straightforward to show based on Eq.~(\ref{eq:time_rev_transformation}) that for each eigenstate of the time-step operator \encapsulateMath{$\hat{W}$} with flux $\phi$, there is a corresponding eigenstate of \encapsulateMath{$\hat{W}(-\phi)$} at the same quasienergy, and the two are related by the time-reversal operator.

Fourth, the last result implies that
the quasienergy bands and band gaps are invariant under inversion of the magnetic flux over the whole lattice.
In addition, by inverting the magnetic flux, the net number of edge modes crossing a given quasienergy contained in a band gap is unchanged.
However, the propagation direction of these edge modes is reversed by the time-reversal operator, since this operator inverts the sign of quasimomentum.

Hence, since the RLBL gap invariant is nothing less than the net number of edge modes associated with a given edge, its sign is reversed if the magnetic flux is reversed over the whole lattice. 
Since the Chern number, $C$, of any set of bands is equal to the difference of RLBL invariants above and below that set of bands, it directly follows that $C(-\phi) = -C(\phi)$.

\section{Floquet topological invariants from the spectral flow}
\label{sec:spectral_flow}
\begin{figure}[t]
	\includegraphics[width=0.99\columnwidth]{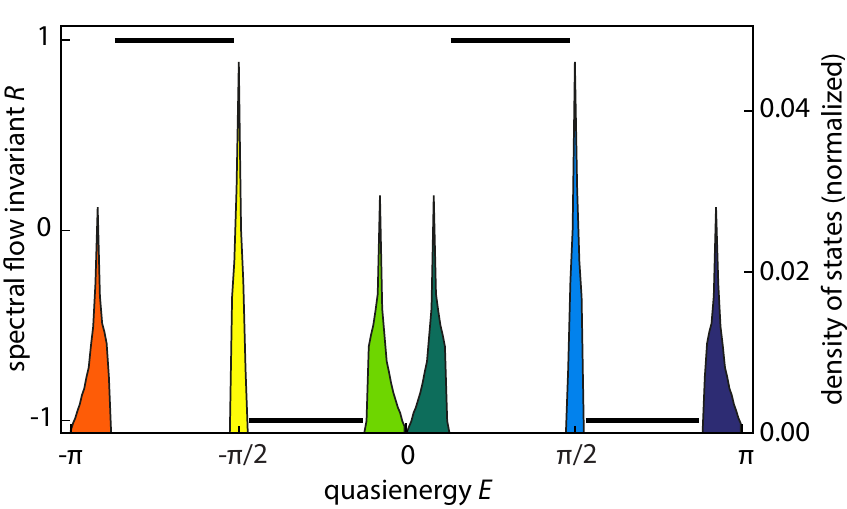}
	\caption{Spectral flow $R$ calculated \cite{Asboth2017} as a function of quasienergy $E$ for a magnetic quantum walk with magnetic flux $\phi=1/3$.
   Using Eq.~(\ref{eq:rudner_number}), we obtain the spectral flow induced by a small change of an additional, fictitious magnetic field, $\beta: 0\rightarrow 2\pi/15$, assuming a fixed quasimomentum $k_x=k_y=1/10$, since the result does not depenend on the chosen quasimomenutm.
  The density of states, normalized to the total number of states, is overlaid (colored areas) to indicate 
  the regions of bulk gaps and quasienergy bands.
  The spectral flow is only defined in the band gaps (thick black lines), where it yields the (integer) RLBL topological invariant, $R$.
   }
	\label{fig:spectral_flow}
\end{figure}

We present a method to calculate the Floquet topological invariant $R$, which we have named RLBL after Ref.~\cite{Rudner2013}.
Instead of evaluating the rather involved three-dimensional winding number of the ``periodized'' time-step operator (winding along quasimomenta $k_x$, $k_y$, and time t), we here apply the method proposed in Ref.~\cite{Asboth2017}.
This method relates the Floquet topological invariant of a certain quasienergy gap to the spectral flow induced through the same gap by a fictitious magnetic field $\beta$, which in the case of magnetic quantum walks is added on top of the magnetic field $B$.
Importantly, the vector potential corresponding to the fictitious magnetic field does not vary inside a magnetic unit cells.

The magnetic field operator \encapsulateMath{$\hat{F}$} is therefore modified to account for the additional, fictitious magnetic field, $\beta$:
\begin{equation}
\hat{F}_1(\beta,B) = \exp[i\hat{\sigma_z}(\beta \lfloor \hat{x}/q \rfloor + B \hat{x})].
\end{equation}
where $\lfloor x/q \rfloor$ is the greatest integer less than or
equal to $x/q$, indexing the magnetic unit cells.
The additional field takes rational values, $\beta=2\pi\,r/s$, with $r$ and $s$ coprime
integers.
The magnetic unit cell is therefore enlarged in the $x$-direction, so as to contain $s\,q\times 1$ lattice plaquettes.
In reality, for the purpose of calculating $R$, as shown below in Eq.~(\ref{eq:rudner_number}), it is sufficient to consider only $\beta=0$ and $\beta =
2\pi/s$.

Since the direct application of the method in Ref.~\cite{Asboth2017} yields the RLBL invariant, $R$, for the quasienergy gap at $E=\pi$, in order to apply the same method for any arbitrary gap containing the quasienergy $\tilde{E}$, we add to the time-step operator an extra term, $\exp(i\tilde{E})$, which shifts the quasienergy spectrum by $-\tilde{E}$, modulo the Floquet zone.
This is equivalent to redefine the branch cut of the complex logarithm; see Sec.~\ref{subsection:bulk_spectrum}.

Thus, the time-step operator of the modified magnetic quantum walk reads as
\begin{equation}
\hat{W}_1(\beta,B,\tilde E) = e^{i\tilde{E}} \,\hat{F}_1(\beta,B)
 \, \hat{S}_y \, \hat{C} \, \hat{S}_x \, \hat{C}.
\end{equation}
Denoting the eigenvalues of \encapsulateMath{$\hat{W}_1$} by $\exp[-i E_j(\beta,B,\tilde{E})]$, with $j = 1, \ldots, 2sq$,
the Floquet topological invariant associated with the gap comprising the quasienergy $\tilde E$ can be simply calculated as:
\begin{align}
	\label{eq:rudner_number}
R &= \frac{1}{2\pi} \left( \sum_{j=1}^{2sq} E_j(1/s,B,\tilde{E}) - \sum_{j=1}^{2sq} E_j(0,B,\tilde{E}) \right).
\end{align}

Figure~\ref{fig:spectral_flow} shows the values of $R$ obtained with the formula in Eq.~(\ref{eq:rudner_number}), plotted as a function of quasienergy $E$.
The figure also reports the density of states, indicating the regions of the bulk gaps and quasienergy bands.
Our results showed that for a quasienergy in a bulk gap, the corresponding invariant is independent of $k_x$ and $k_y$, and agrees with
the number of edge states seen in the paper.

\section{Motional excitations}
\label{app:perturbative_force}
We derive the perturbative potential, $\hat{H}_\phi$, which is shown in Eq.~(\ref{eq:hamilt_force}), acting onto the walker when a spin-dependent linear potential gradient is switched on and off stroboscopically.
According to Sec.~\ref{sec:artificialgaugefieldimp}, an artificial flux $\phi$ can be produced by flashing for a time $\tau$ the spin-dependent potential gradient:
\begin{equation}
	\label{eq:pertur_pot_exact}
	\hat{H}_\phi = \frac{\hbar\hspace{0.5pt}}{\tau}B\hspace{0.5pt} \hat{x}\hspace{0.5pt} \hat{\sigma}_z\hspace{0.5pt},
\end{equation}
where $B=2\pi\phi$ is the magnetic field, and $x\in \mathbb{Z}$ is the lattice site coordinate in the $x$-direction.

In the limit of a deep optical lattice, $V_0\gg E_R$, the potential around a given lattice site can be approximated by a harmonic potential with harmonic trap period $\tau_\text{HO} = \sqrt{m \lambda_L^2 /V_0}$; see Sec.~\ref{sec:two_dim_quantum_walks}. Hence, by expressing the position operator in Eq~(\ref{eq:pertur_pot_exact}) in terms of the operators creating and annihilating motional excitations,
\begin{equation}
	\hat{x}=\frac{1}{\pi(8\hspace{0.5pt}V_0/E_R)^{1/4}}(\hat{b}+\hat{b}^\dagger)\hspace{0.3pt},
\end{equation}
one can directly obtain Eq.~(\ref{eq:hamilt_force}).

\bibliographystyle{apsrev4-1}
\bibliography{magnetic_QW}

\begin{thebibliography}{102}
\bibitem [{sca()}]{scattering_prob}%
  \BibitemOpen
  \bibinfo {note} {Since for a given imprinted phase, the
  product of the intensity and pulse duration is fixed, the probability of
  scattering a photon, which depends on the same product, is fixed as
  well}\BibitemShut {NoStop}%
\end{thebibliography}

\end{document}